\documentclass[fleqn,10pt]{wlscirep}
\usepackage[utf8]{inputenc}
\usepackage[T1]{fontenc}
\usepackage{lineno}

\title{Visible dual-comb spectroscopy across more than 100~THz with lithium niobate nanophotonic waveguides}

\author[1,2,†]{Carter Mashburn}
\author[3,†]{Kristina F. Chang}
\author[1,†]{Michael J. Wahl}
\author[4]{Mathieu Walsh}
\author[1]{Daniel I. Herman}
\author[1,2]{Matthew Heyrich}
\author[1]{Tsung-Han Wu}
\author[3]{Nazanin Hoghooghi}
\author[5]{Ryoto Sekine}
\author[5]{Luis Ledezma}
\author[1]{Emily Jerris}
\author[5]{Alireza Marandi}
\author[4]{Jérôme Genest}
\author[1,2,*]{Scott A. Diddams}

\affil[1]{Electrical, Computer, and Energy Engineering, University of Colorado Boulder, Boulder, CO 80309, USA.}
\affil[2]{Department of Physics, University of Colorado Boulder,  Boulder, CO 80309, USA.}
\affil[3]{Time and Frequency Division, National Institute of Standards and Technology, Boulder, CO 80305, USA.}
\affil[4]{Department of Electrical and Computer Engineering, Université Laval, Québec, QC G1V 0A6, Canada.}
\affil[5]{Department of Electrical Engineering, California Institute of Technology, Pasadena, CA 91125.}
\affil[*]{Corresponding author: scott.diddams@colorado.edu, \textsuperscript{†}These authors contributed equally.}

\begin{abstract}

Broadband and high-resolution spectroscopy in the visible and ultraviolet is central to advances in multiple fields, including fundamental quantum physics, biology, atmospheric science and astronomy. Traditionally, these measurements are performed with grating or Fourier-transform spectrometers using incoherent light sources. Leveraging coherent light enables powerful frequency-comb-based techniques, but is limited by the technical complexity of efficiently generating broad spectral bandwidths from relatively narrowband and spectrally distant laser sources. Current visible dual-comb spectrometers require implicit compromises between optical bandwidth, experimental simplicity, and acquisition speed. In this work, we introduce a simple and efficient dual-comb spectrometer that converts robust Er:fiber frequency combs from the near-infrared to the ultraviolet and visible with thin-film lithium niobate (TFLN) nanophotonic waveguides. Using real-time signal processing, we retrieve coherently averaged dual-comb spectra over nearly 120 THz of simultaneous bandwidth in the visible with 100~MHz spectral resolution. With these capabilities, we measure the broadband absorption spectrum of molecular iodine (I\textsubscript{2}), demonstrating the broadest visible spectral coverage of a dual-comb spectrometer to date. Additional measurements of NO\textsubscript{2}, atomic rubidium, and atomic sodium further illustrate the achievable combination of spectroscopic bandwidth, resolution, and intrinsic frequency accuracy. 
Our results demonstrate the powerful integration of low-power frequency combs, nonlinear nanophotonics, and digital signal processing to enable a compact, efficient and versatile approach to high-resolution mapping of complex absorption spectra across 500~THz in the UV-visible and near-infrared spectral regions for multiple applications beyond the research lab.

\end{abstract}
\begin{document}

\flushbottom
\maketitle

\thispagestyle{empty}

Absorption spectroscopy in the ultraviolet (UV) and visible (VIS) spectral regions is a ubiquitous tool for mapping the electronic structure of atoms, ions, and molecules. Coherently accessing this information is key to enabling important applications such as optical clocks and timekeeping\cite{RevModPhys.87.637,Fortier:26}, the measurement of reaction dynamics in physical chemistry and plasma physics\cite{Ndengu2021,Adamovich_2022}, astronomical spectrograph calibration\cite{Osterman2007,Wilken2012,Glenday2015} and atmospheric sciences\cite{Galiter:2020}. Still, a lack of broadband (>100~THz) coherent light sources and high-resolution (<1~GHz) spectroscopic instrumentation in the ultraviolet and visible spectral regions is limiting advances in these areas. For example, simultaneously resolving and mapping the absolute UV-VIS frequencies of thousands of transitions in complex molecular spectra is important to advancing fundamental physical chemistry beyond the Born-Oppenheimer approximation\cite{Bera2008,Ndengu2023}. In addition, coherent broadband UV-VIS sources are necessary to track multi-species dynamics in complex plasma systems\cite{Cunge_2011}. Finally, measuring the absorption spectrum of molecular species used for astronomical spectrograph calibration with light sources directly linked to absolute frequency standards will impact exoplanet studies and searches for habitable planets outside our solar system\cite{Schmidt:Espresso,Schmidt2021,Reiners2024}.\par

\begin{figure*}[ht!]
\centering
\includegraphics[width=5 in]{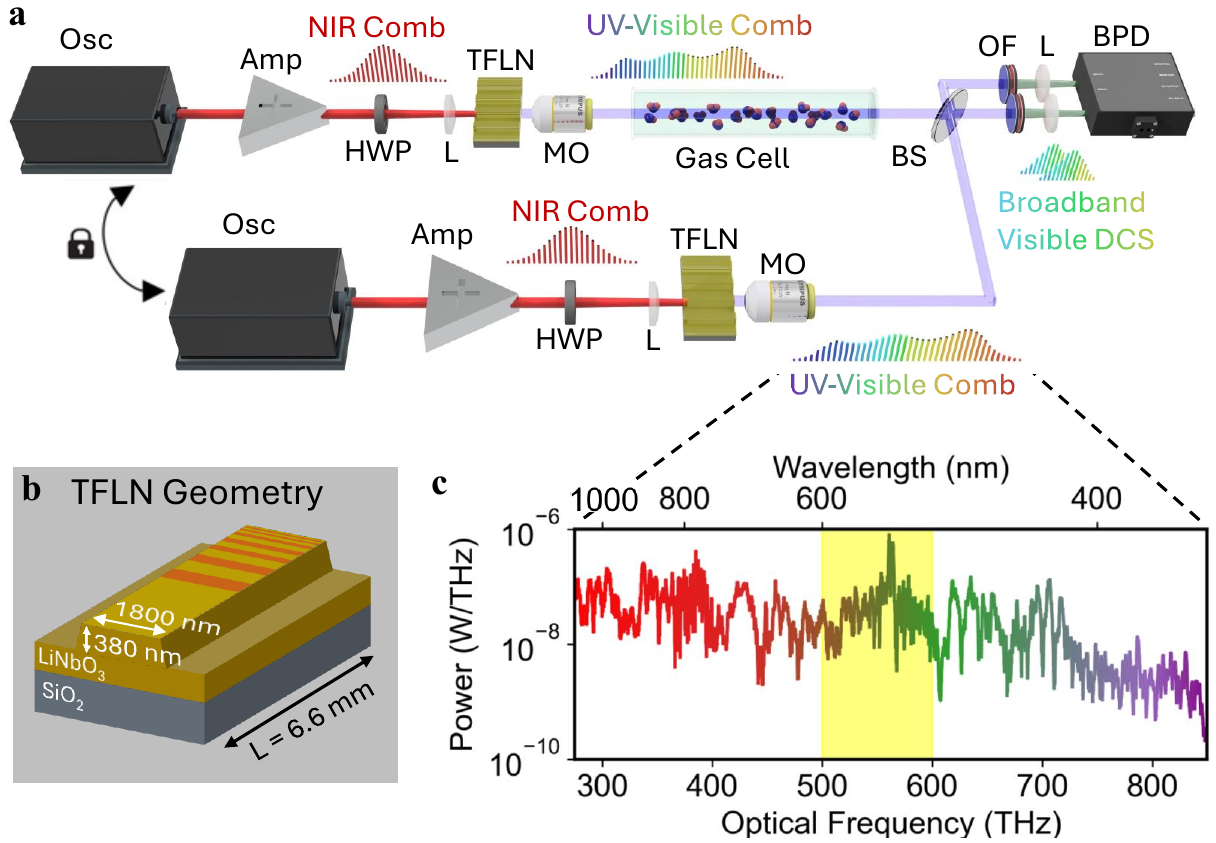}
\caption{Experimental setup for dual-comb spectroscopy (a) Two frequency combs independently generate broadband UV-VIS spectra in thin-film lithium niobate nanophotonic waveguides for dual-comb spectroscopy. (b) Schematic of the thin-film lithium niobate waveguide with chirped poling. (c) Generated UV-visible-NIR spectrum and the isolated region utilized for dual-comb measurements of iodine (yellow region). Summary of abbreviations in panel (a): Osc: Er:fiber oscillator; Amp: Er:fiber amplifier; HWP: half-wave plate; L: lens; TFLN: thin-film lithium niobate; MO: microscope objective; BS: beamsplitter; OF: optical filters; PD: photodetector.}
\label{fig:1}
\end{figure*}

Dual-comb spectroscopy (DCS) using coherent optical frequency combs provides a powerful spectroscopic technique capable of interrogating narrow-linewidth transitions in atoms and molecules over broad optical bandwidths with fast acquisition speeds. Currently, DCS has been predominantly applied to measuring rovibrational transitions in the near-infrared (NIR) and mid-infrared (MIR) where frequency comb technology is more mature\cite{Okubo_2015,Ycas2018,Lind2020,Hoghooghi:24,Herman2021,Guay:18,Guay:19,Muraviev2020,Abbas2019,Yu2018,KONNOV2025}. Pushing frequency combs and DCS to the visible and ultraviolet spectral regions for the mapping of electronic transitions has remained a challenge. Central to this challenge is the straightforward and low-power generation of frequency combs with sufficient spectral bandwidth to cover entire absorption bands that can span hundreds of terahertz in this spectral region\cite{Schuster:2021}.\par

Nonlinear optics plays an essential role in broadband UV-VIS comb development, as there are currently no suitable gain media for native comb generation. Consequently, the leading strategy is to upconvert NIR and MIR frequency combs to the UV-VIS using cascaded nonlinear frequency mixing in bulk materials, fibers, and waveguides\cite{Chang:2024,Eber:2024,Di:2023,Xu:2024,Stroud2025,Hofer2025,Li2025,fuerst2025,McCauley:2024,Muraviev2024,Sugiyama:2023,Ideguchi2012,Eber:25,Tian2024,Pal2025,Furst:2024,Weeks2022}. While such techniques have achieved instantaneous UV-VIS bandwidths approaching 100~THz\cite{Muraviev2024,Kirchner:25}, the comparatively low conversion efficiencies typically require lasers with multi-watt average powers and result in bulky, expensive, and complex setups for UV-VIS frequency comb generation.\par


In this work, we address this challenge by pumping thin-film lithium niobate (TFLN) nanophotonic waveguides with low-power 1550~nm fiber lasers to efficiently drive broadband (500~THz) UV-VIS-NIR frequency comb generation. The sub-micron mode confinement over extended interaction lengths enabled by nanophotonics allows for efficient nonlinear mixing with low on-chip pulse energies ($\sim$100~pJ). Furthermore, the nanophotonic platform can be dispersion-engineered through choice of waveguide geometry, and, in the case of TFLN, can be spectrally tailored through quasi-phase matching with periodic-poling. As described in our previous work\cite{Wu:2024}, cascaded quasi-phase matching permits the generation of frequency combs over ultra-broad bandwidths that are otherwise difficult to achieve with conventional bulk nonlinear optical crystals and devices. Ultimately, the reduction in laser power requirement combined with chip-based cascaded harmonic generation achieves a low SWaP (size, weight and power) form factor for UV-VIS generation, making this source a promising sub-system for portable optical spectroscopy and future integrated sensing platforms.\par

We demonstrate the capabilities of TFLN devices in DCS by measuring
the broadly absorbing B-X transition in molecular iodine (I\textsubscript{2}) in the visible region, covering approximately 500~nm-600~nm. In this spectral range, our spectrometer resolves thousands of complex absorption features with 100~MHz resolution in a single measurement. Our measured spectrum spans nearly 120~THz: the broadest demonstration of DCS in the visible spectral region to date. Furthermore, we perform precision spectroscopy of the 5S-5P and 3S-3P transitions in atomic rubidium and sodium, respectively, at multiple spectral harmonics. Together, the results and capabilities of this system highlight a powerful approach to broadband, high-resolution NIR, visible and UV spectroscopy in a compact package.



\begin{figure*}[ht]
\centering
\includegraphics[width=16 cm]{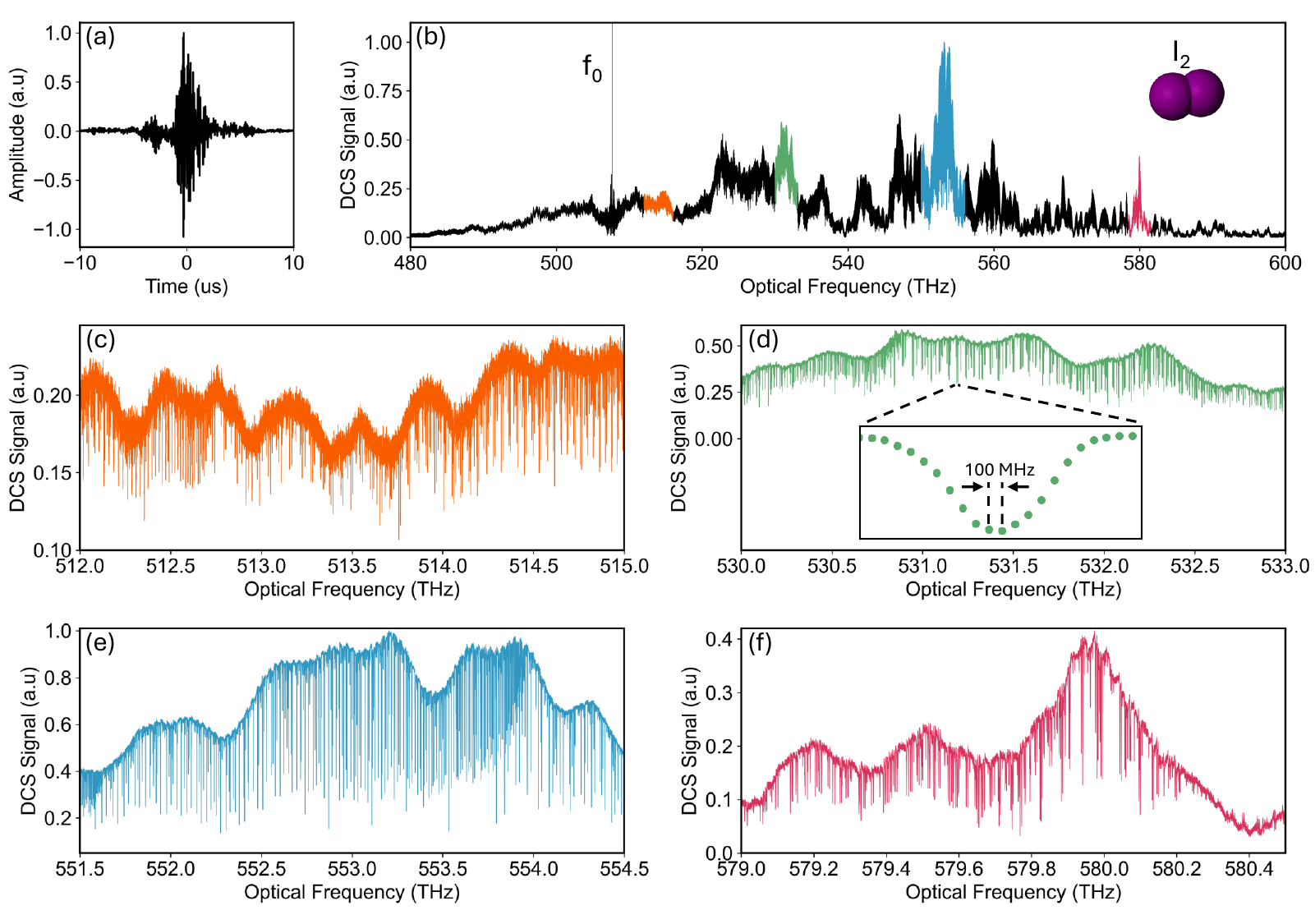}
\caption{Dual-comb interferogram and spectrum from iodine spectroscopy. (a) Time-averaged interferogram. (b) Corresponding dual-comb spectrum with $f_0$ tone present. The black curve shows the entire spectrum, while the colored regions are expanded in (c)-(f). The inset in (d) shows a single absorption feature and the 100~MHz point spacing.}
\label{fig:2}
\end{figure*}

\section*{Experiment}
The DCS system (Fig. \ref{fig:1}(a)) utilizes TFLN waveguides to generate broadband UV-VIS-NIR frequency combs with over 500~THz of bandwidth spanning 350-1000~nm. Conversion efficiency to ultraviolet-visible wavelengths as high as 17\%\cite{Wu:2024} relaxes the on-chip pulse energy requirements to the 100 pJ level, such that we can use simple front-end lasers based on compact and robust Er:fiber oscillators operating near $f_r=100$~MHz. Similar Er:fiber combs have been packaged and deployed for a variety of spectroscopic and timing applications beyond the research lab\cite{Yun:23, MenloSoundingRocket, Roslund2024,Sinclair:14}. While we use tabletop components here, the experimental concept for chip-based UV-VIS generation and spectroscopy presented in this work could be fully packaged in a compact system for use in wide-ranging environments.

The two combs are offset-frequency stabilized and optically phase-locked to a common narrow-linewidth (<10~Hz) continuous-wave (CW) laser. They are amplified to produce a modest 300~mW of average power with pulse durations of approximately $\sim$100~fs and a central wavelength near 1550~nm.  These NIR pulses are coupled into free-space and pass through a half-wave plate before coupling to the TFLN waveguides with an aspheric lens. The estimated coupling loss is 10~dB at each facet of the waveguide, such that 30~mW is coupled into the waveguide when pumped with 300~mW. The inclusion of optimized broad bandwidth inverse tapers at the input facet could significantly reduce the input coupling loss to <0.5 dB\cite{Chen2024}, and low-loss output coupling across hundreds of terahertz remains largely unexplored. 

Broadband UV-VIS-NIR comb light is coupled off-chip with a microscope objective, providing roughly 1.2~mW of optical power spanning 350-1000~nm. One comb is subsequently sent through a gas cell before being spatially overlapped with the other comb on a beamsplitter. The co-propagating combs pass through a trio of optical filters to isolate the 500~nm-600~nm region of interest: one longpass filter which cuts near 450~nm and two shortpass filters which cut near 650~nm and 750~nm. Subsequent dual-comb detection is performed with a repetition rate difference near $\Delta f_r$=33~Hz on a balanced silicon photodetector. Additional technical details on the experimental setup are found in the Supplement.\par

\begin{figure*}[ht]
\centering
\includegraphics[width=18cm]{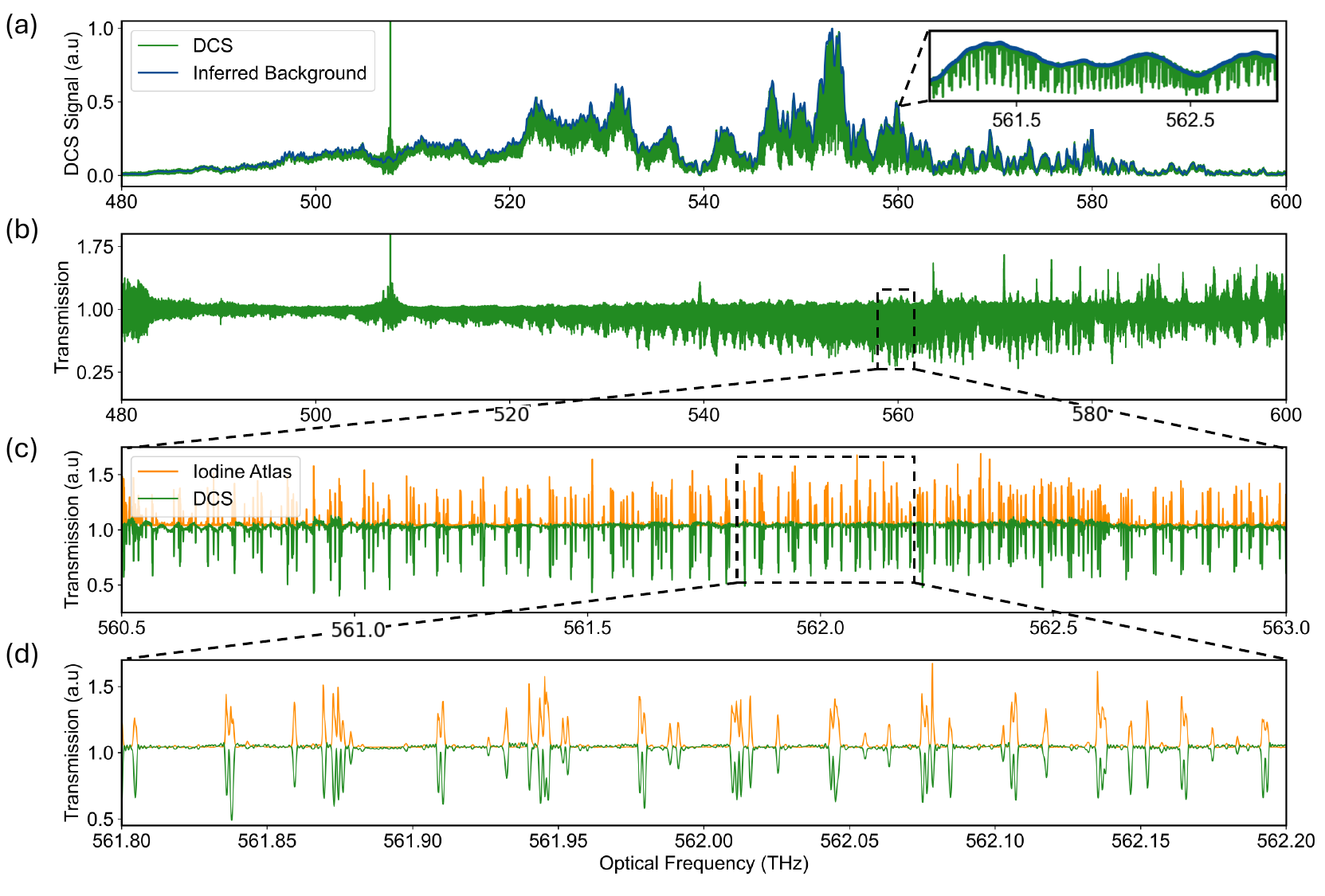}
\caption{Processed iodine dual-comb spectrum and comparison to iodine atlas. (a) Measured dual comb spectrum (green) and inferred background obtained through cepstral analysis (blue). Inset shows an expanded region for comparison. The spectral artifact near 507~THz is an $f_0$ tone. (b) Baseline-corrected transmission spectrum. (c) Comparison of baseline-corrected transmission spectrum to iodine atlas\cite{SALAMI&ROSS}. The iodine atlas has been inverted, scaled, and vertically shifted to aid comparison. (d) Expanded region of (c), showing agreement in positions of absorption features between our measurement and the iodine atlas.}
\label{fig:3}
\end{figure*}
\subsection*{Waveguide Geometry and Spectra}
The dimensions and geometry of the TFLN waveguides are shown in Fig. \ref{fig:1}(b). The design features two distinct sections along the longitudinal direction: an initial 3~mm un-poled segment where the input pulse is spectrally broadened via $\chi^{(3)}$ self-phase modulation, followed by a 3.6~mm poled segment with a chirped poling period that linearly decreases from 12.5~$\mu$m to 2.5~$\mu$m. In the chirped poling region, quasi-phase matching enables the cascaded $\chi^{(2)}$ generation of broadband second, third, and fourth harmonics of the driving laser in the NIR, visible, and UV, respectively. In our dual-comb system, the two TFLN waveguides differ in their length, cladding and insertion loss. For clarity we report characteristics of only one of the waveguide geometries, with additional details on the other waveguide found in the Supplement. Further details on the fabrication of the waveguides and their properties can be found in Ref. \cite{Wu:2024}.\par

A UV-VIS-NIR spectrum produced by the TFLN waveguides with approximately 30 mW of waveguide-coupled pump power is shown in Fig. \ref{fig:1}(c). As seen, the spectrum spans the full visible and extends into the ultraviolet, with the spectrally filtered region used for iodine DCS highlighted in yellow. A present challenge for spectroscopy is the modulated structure, which makes it difficult to produce two combs with identical overlapping spectra. Using mis-matched spectra in DCS limits the signal-to-noise ratio (SNR) of measurements\cite{Coddington:2016}. Additionally, in the region where we perform dual-comb spectroscopy, the integrated power is on the order of $\sim$10~$\mu$W corresponding to approximately $\sim$10~pW per comb mode. The optical power per comb mode is the primary limitation of the SNR in the current system.\par

\begin{figure*}[ht!]
\begin{centering}
\includegraphics[width=6 in]{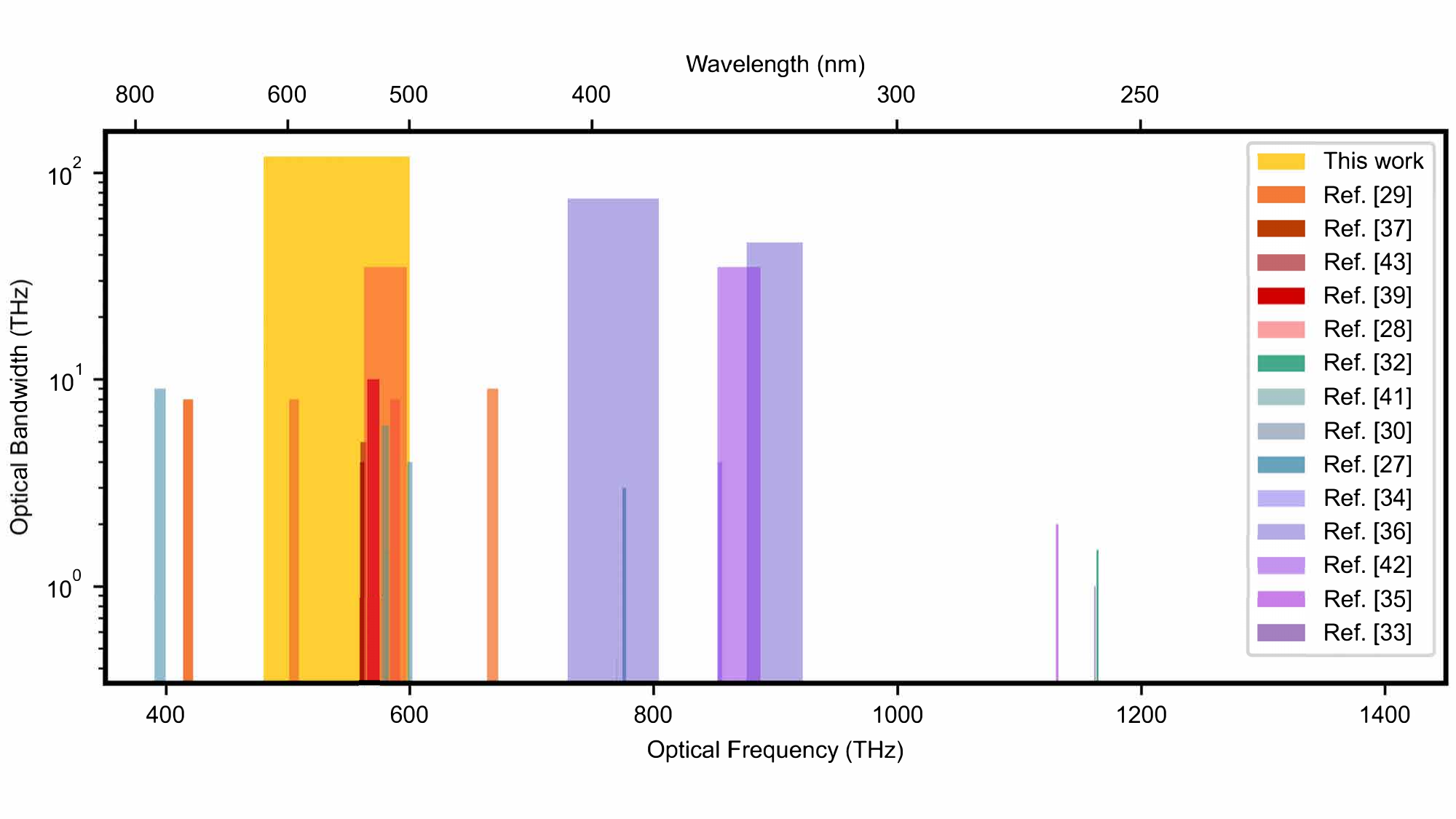}
\caption{Comparison of optical bandwidths of recent dual-comb works in the ultraviolet and visible spectral ranges}
\label{fig:4}
\end{centering}
\end{figure*}

The spectral power density can be increased with improved waveguide coupling, as noted above. Additionally, the generation of flatter spectra, as has been accomplished with $\chi^{(3)}$\cite{Genier2021,Carlson2018} supercontinuum and difference-frequency-generation in $\chi^{(2)}$\cite{Huber2000,Kowalczyk2023,Timmers:18} materials, is still a topic of research for combs built from spectrally broadened harmonics~\cite{Fan:24}. Some of the structure is expected from the overlapping and interfering harmonics\cite{fanPRL}, but this could be partially addressed by spectrally multiplexing the harmonics on separate detectors, as in our previous work\cite{Chang:2024}. Future work will further address the efficiency and flatness through improved waveguide designs, and some preliminary models have been introduced in our earlier work\cite{Wu:2024}. Within this complex design space, we anticipate that improved coupling efficiency will allow for a wider range of waveguide designs, including shorter waveguides that minimize cumulative interference effects that follow the breakup of the pump pulse.

\subsection*{Data Acquisition and Processing}
Real-time processing is a practical requirement for the long measurement durations needed to achieve high SNR over the broad optical bandwidths generated in this system. Digitizing interferograms and referencing beat notes at the repetition rate (100~MHz) with high dynamic ranges (16 bits per sample) leads to an 800 MB/s data stream that would fill 1~TB of storage in $\sim$20 minutes. To mitigate this significant storage and post-processing overhead, our acquisition scheme utilizes computation parallelization on a graphics processing unit (GPU) for real-time coherent averaging \cite{WalshGPU}.\par

The real-time GPU-based algorithm first performs phase-corrections\cite{Ycas2018,Roy:12} using phase-noise information extracted from the sampled $f_0$ and $f_{opt}$ of each comb to overcome fluctuations on fast time-scales. Self-corrections\cite{Hebert:17,Hebert:18} are then performed to overcome slow out-of-loop fluctuations not captured by the referencing signals. The algorithm then aligns all the interferograms on a common sampling grid and averages them for a user-defined period. For our optical and referencing configuration, the real-time averaging enabled a storage reduction of over four orders of magnitude in comparison to a raw stream of interferograms. Technical details regarding the operation of this data acquisition system can be found in Ref. \cite{WalshGPU} and the Supplement.\par



\section*{Results and Discussion}
\subsection*{High-Resolution Dual-Comb Spectroscopy of Iodine}
Using the filtered region from 500~nm-600~nm, we interrogate the dense B-X rovibronic transitions of molecular iodine with 100~MHz resolution.
Iodine was chosen due to the availability of high-resolution transmission spectra over broad bandwidths in the Doppler-limited regime\cite{SALAMI&ROSS}. The absorption spectrum of iodine also serves as an important tool for the calibration of astronomical spectrographs and other spectral measurements. Broadband measurements of the iodine absorption spectrum with DCS provides direct traceability to absolute frequency standards, which could potentially replace reference spectra taken with Fourier-transform spectroscopy (FTS)\cite{Schmidt:Espresso,Schmidt2021,Reiners2024}.\par

The time-averaged DCS interferogram of iodine and its corresponding spectrum are shown in Fig. \ref{fig:2}(a) and (b), respectively, and were obtained by averaging approximately 200,000 interferograms over the course of 115~min. The dual-comb spectrum spans nearly 120~THz, defined as the bandwidth with which our dual-comb SNR is above unity (see Supplement). Despite the highly modulated structure of the spectrum produced by the TFLN waveguides, we still measure iodine absorption features across the dual-comb spectrum with 100~MHz resolution, as shown in the expanded colored regions in Fig.  \ref{fig:2}(c)-(f).\par

A consequence of the large optical bandwidths covered by our system is that the spectral harmonics of our 1550~nm driving laser overlap, producing an $f_0$ RF tone in the photodetector signal. This is revealed as an artifact (labeled in Fig. \ref{fig:2}(b) near 507 THz) when the dual-comb RF spectrum is scaled back to the optical domain. The generation of $f_0$ tones in frequency combs which are generated through cascaded $\chi^{(2)}$ processes has been studied \cite{LindPRL,fanPRL}. While we observe no noticeable contribution to our spectroscopy from the overlapping harmonics in this experiment, future work will aim to further understand the impact of such interleaved comb structures in DCS measurements.\par

To compare against existing iodine transmission data, we use cepstral analysis to infer the background intensity spectrum of our dual-comb measurement, shown in blue in Fig. \ref{fig:3}(a)\cite{Cole:Cepstral}. Removal of the background results in the transmission spectrum shown in Fig. \ref{fig:3}(b). The areas where the transmission is significantly greater than unity correspond to those areas in the dual-comb spectrum where no signal was measured and can therefore be disregarded. One region is shown in Fig. \ref{fig:3}(c) and is compared against a broadband iodine atlas in this spectral region with a resolution of approximately 600~MHz\cite{SALAMI&ROSS}. Fig. \ref{fig:3}(d) provides an expanded region of this comparison. Cross-correlating the central 100~THz of our data (from $\approx$490-590~THz) with the iodine atlas over the same region shows agreement within the stated 90~MHz uncertainty of the atlas.


However, we find discrepancies in the measured absorption depths. More detailed study is needed to understand the source of this disagreement, but we note that such spectral amplitude discrepancies are common to other iodine absorption data from current literature in the 500~nm-600~nm spectral region \cite{FERNANDEZ2023,Torres2022,Sugiyama:2023}. Our TFLN DCS system could be a powerful tool to quantitatively resolve such questions, but this will require improved SNR and better knowledge of the spectral baseline. Higher SNR can be achieved through increasing the spectral power per-comb-mode and will require improvements in waveguide coupling and design, as previously mentioned. Improving our knowledge of the spectral baseline will require adjusting the optical configuration and detection scheme to permit the simultaneous acquisition of a spectral baseline \cite{Newbury:sensitivity,Coddington:2016}.


To put our results in context, we compare the broad DCS bandwidths afforded by our TFLN devices to other recent dual-comb measurements in the ultraviolet and visible spectral regions\cite{Chang:2024,Eber:2024,Di:2023,Xu:2024,Stroud2025,Hofer2025,Li2025,fuerst2025,McCauley:2024,Muraviev2024,Sugiyama:2023,Eber:25,Pal2025,Furst:2024,Weeks2022} (Fig. \ref{fig:5}). Only a few of these works achieve bandwidths exceeding 10~THz, and they do so using driving lasers with pulse energies on the order of 10's of nanojoules (several watts of optical power at 100 MHz). Our work stands alone in this spectral region as the only one to surpass 100~THz of simultaneous bandwidth with only 300 pJ of on-chip pulse energy, further highlighting the benefits of TFLN nanophotonics in efficiently generating broadband visible frequency combs. Even still, we have nearly 400~THz of unused bandwidth in the UV, visible and NIR. With improvements to waveguide insertion loss and the addition of spectral multiplexing\cite{Chang:2024}, the entire bandwidth could be employed for DCS to achieve spectral coverage comparable to grating-based spectrometers and FTS, but with superior resolution and fast acquisition times.\par

The same bandwidth and resolution can also be applied to more complex and atmospherically relevant molecules, such as ozone ($\rm{O_3})$, formaldehyde ($\rm{H_2CO}$), and nitrogen dioxide ($\rm{NO_2})$. This trio of molecules is particularly important in the troposphere, where complex photochemistry links them together through nonlinear chemical cycles. Nitrogen dioxide and formaldehyde, for example, are the photolytic precursors to the formation of ozone and various other aerosols and pollutants that can negatively impact human health and crop growth\cite{Huan:CleanAir,Lobell:CropGrowth}. To understand the complex interplay between $\rm{O_3}$, $\rm{H_2CO}$, $\rm{NO_2}$ and other atmospheric species and pollutants, broadband and high-resolution measurements of their UV-VIS-NIR spectrum are desirable, and we present preliminary measurements of nitrogen dioxide at a reduced resolution in the Supplement. Similar high-resolution and broadband measurements of ozone would also be impactful in the 350~nm-1000~nm\cite{ozone_spectroscopy}. Lastly, employing the full spectral bandwidth generated by the TFLN waveguides for DCS could enable the simultaneous and real-time monitoring of the highly coupled concentrations of ozone, formaldehyde, and nitrogen dioxide\cite{ozone_production}.

\par

\subsection*{Dual-Comb Spectroscopy of Atomic Rubidium and Sodium}
\begin{figure*}[ht!]
\begin{centering}
\includegraphics[width=6 in]{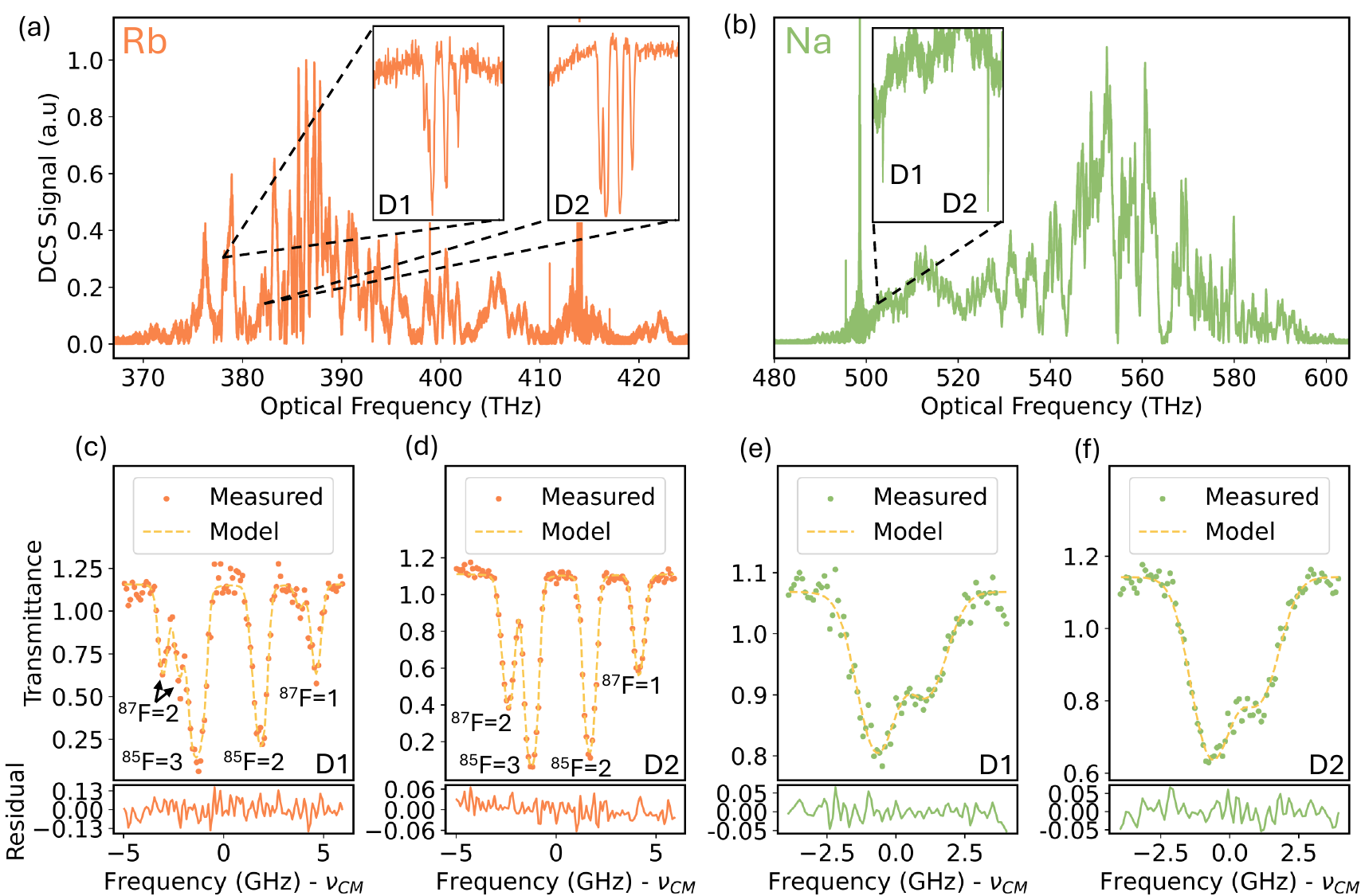}
\caption{Dual comb spectroscopy of atomic rubidium and sodium. Broadband dual-comb spectrum with coverage of (a) atomic rubidium and (b) atomic sodium. The insets shows a zoom-in of the measured transitions. (c)-(d) Measured rubidium D1 and D2 transitions after cepstral analysis (orange) compared to the model (yellow). Transition labels denote the respective ground-hyperfine state and isotope. Residuals (Measurement-Model) are shown in the bottom panel. For the D1 transitions, $\nu_{CM}$ = 377,107.409~GHz, and for the D2 transitions $\nu_{CM}$ = 384,230.436~GHz. The D1 and D2 transitions were measured at two different temperatures, with details found in the Supplement. (e)-(f) Measured sodium D1 and D2 transitions after cepstral analysis (green) compared to the model (yellow). Residuals (Model-Measurment) are shown in the bottom panel. For D1 $\nu_{CM}$ = 508,333.152~GHz, for D2 $\nu_{CM}$ = 508,848.718~GHz.}
\label{fig:5}
\end{centering}
\end{figure*}
We further illustrate the capabilities of our spectrometer to extract high-resolution spectra from spans approaching 100 THz with precision spectroscopy of the $5S-5P$ and $3S-3P$ transitions in atomic rubidium and sodium vapors near 380~THz and 508~THz, respectively. To change the spectral interrogation region, we need only change the spectral filters used before detection of the dual-comb interference signal. However, the addition of spectral multiplexing could allow for the simultaneous interrogation of multiple transitions in multiple species in the future, which is increasingly desirable for use in atom-based searches for physics beyond the Standard Model\cite{DCS_Budker, RevModPhys_Safronova}.

The measured dual-comb spectrum of atomic rubidium is shown in Fig.\ref{fig:5}(a) and was obtained from 30 minutes of averaging. In Fig.\ref{fig:5}(c) and (d), we show the measured 5\textsuperscript{2}S\textsubscript{1/2}-5\textsuperscript{2}P\textsubscript{1/2} (D1) and 5\textsuperscript{2}S\textsubscript{1/2}-5\textsuperscript{2}P\textsubscript{3/2} (D2) lines and fit a model based on their well-known hyperfine structure. Details on the models used are found in the Supplement. We observe contributions from \textsuperscript{85}Rb and \textsuperscript{87}Rb and resolve their ground-state hyperfine splittings of 3~GHz and 6.8~GHz, respectively. The multiple excited-state splittings are not fully resolved due to their proximity and $\sim$500 MHz (FWHM) of Doppler-broadening.\par

Accounting for the known hyperfine transition frequencies in both isotopes, along with the relative line strengths and natural isotopic abundance\cite{Siddons2008}, our DCS measurements agree with previous measurements of the center-of-mass transition frequencies at the <10~MHz level \cite{Siddons2008}. The uncertainty in this discrepancy is statistical in nature and is dominated by the SNR and imperfect knowledge of the spectral baseline. Additional details on this analysis can be found in the Supplement.
\par 

The measured dual-comb spectrum of atomic sodium is shown in Fig.\ref{fig:5}(b) and was obtained from 80 minutes of averaging. In Fig.\ref{fig:5}(e) and (f), we show the measured 3\textsuperscript{2}S\textsubscript{1/2}-3\textsuperscript{2}P\textsubscript{1/2} (D1) and 3\textsuperscript{2}S\textsubscript{1/2}-3\textsuperscript{2}P\textsubscript{3/2} (D2) lines and corresponding model fits (see Supplement). Here, we observe a deviation from earlier measurements of the line centers that did not employ frequency combs. In particular, our determination of the D1 line center is red-shifted by 43(25)~MHz with respect to widely-cited iodine-referenced dye laser measurements from 1981\cite{juncar1981}. To the best of our knowledge, we provide the first absolute frequency-comb measurements of these transitions. In conjunction with the molecular measurements presented above, our results demonstrate the potential value of our system in expanding and improving existing atomic spectral databases, especially for rare-earth and actinide elements whose atomic spectra are more complex\cite{DCS_Budker}.


\section*{Conclusion and Outlook}
In conclusion, this work demonstrates DCS using broadband visible frequency combs generated by TFLN nanophotonic waveguides. We performed the broadest DCS experiment in the visible region to date, and measured gas-phase iodine spectra in the 500~nm-600 nm region with 100~MHz spectral resolution. Broadband spectroscopy of nitrogen dioxide and precision spectroscopy of atomic rubidium and sodium were also performed. Importantly, the system we employ is built on turnkey, compact dual Er:fiber lasers, highlighting the potential of similar systems to be integrated into low-SWaP packages.

Our results highlight a path towards extending TFLN DCS systems over the full NIR to UV bandwidth spanning over 500~THz. Improvements in fiber-to-chip coupling efficiency together with spectrally multiplexed detection will enable higher SNR measurements with faster acquisition speeds. In addition, TFLN design improvements leading to flatter spectra will also improve the SNR and allow for more rigorous quantitative spectroscopy, further enabling the necessary miniaturization of broadband, high-resolution, and high-speed spectroscopy that will bring lab-grade precision to many impactful applications.


\section*{Data availability}
All data required to reproduce the figures in this paper will be available via the University of Colorado CU Scholar at https://scholar.colorado.edu.

\section*{Code availability}
Codes used to analyze the data in this work may be obtained from the authors upon reasonable request.

\section*{Acknowledgements}
This work was funded by the U.S. Air Force (FA9550-16-1-0016, FA9550-22-1-0483), the NSF QLCI Award OMA-2016244, and the National Institute of Standards and Technology. The authors thank P. Sekhar, C. Fredrick, and D. Meyer for valuable discussions and feedback and \'ElanSpectral for use of their PUG software. Device nanofabrication was performed at the Kavli Nanoscience Institute (KNI) at Caltech.

\section*{Author contributions}
The experiments were performed by C.M, K.F.C, M.J.W, with assistance from E.J, N.H, M.H, and T.-H.W. Data analysis was performed by C.M, M.J.W, M.W and D.I.H. The paper was written by C.M, K.F.C, M.J.W and S.A.D with editing and input from all authors. T.-H.W and S.A.D conceived the waveguide designs, and L.L. and R.S. fabricated the waveguides under the supervision of A.M. M.W and J.G. developed and provided the data acquisition system. S.A.D and J.G. supervised the project.

\section*{Additional Information} 
See the Supplementary document for supporting content.

\section*{Competing financial interests} 
M.W. and J.G. are founders of \'ElanSpectral which produces the data acquisition and phase-correction software. R.S. and A.M. are involved in developing photonic integrated nonlinear circuits at PINC Technologies Inc. R.S., L.L., and A.M. have an equity interest in PINC Technologies Inc. 

\bibliography{refs}

@Article{Sugiyama:2023,
author={Sugiyama, Yohei
and Kashimura, Tsubasa
and Kashimoto, Keiju
and Akamatsu, Daisuke
and Hong, Feng-Lei},
title={Precision dual-comb spectroscopy using wavelength-converted frequency combs with low repetition rates},
journal={Scientific Reports},
year={2023},
month={Feb},
day={13},
volume={13},
number={1},
pages={2549},
issn={2045-2322},
doi={10.1038/s41598-023-29734-2},
}

@article{Eber:2024,
author = {Alexander Eber and Lukas F\"{u}rst and Florian Siegrist and Adrian Kirchner and Benedikt Tschofenig and Robert di Vora and Armin Speletz and Birgitta Bernhardt},
journal = {Opt. Express},
number = {4},
pages = {6575--6586},
publisher = {Optica Publishing Group},
title = {Coherent field sensing of nitrogen dioxide},
volume = {32},
month = {Feb},
year = {2024},
doi = {10.1364/OE.513523},
}

@article{Furst:2024,
author = {Lukas F\"{u}rst and Adrian Kirchner and Alexander Eber and Florian Siegrist and Robert di Vora and Birgitta Bernhardt},
journal = {Optica},
number = {4},
pages = {471--477},
publisher = {Optica Publishing Group},
title = {Broadband near-ultraviolet dual comb spectroscopy},
volume = {11},
month = {Apr},
year = {2024},
doi = {10.1364/OPTICA.516783},
}

@article{Di:2023,
author = {Yuanfeng Di and Zhong Zuo and Daowang Peng and Daping Luo and Chenglin Gu and Wenxue Li},
journal = {Photon. Res.},
number = {8},
pages = {1373--1381},
publisher = {Optica Publishing Group},
title = {Dual-comb spectroscopy from the ultraviolet to mid-infrared region based on high-order harmonic generation},
volume = {11},
month = {Aug},
year = {2023},
doi = {10.1364/PRJ.486864},
}

@Article{Xu:2024,
author={Xu, Bingxin
and Chen, Zaijun
and H{\"a}nsch, Theodor W.
and Picqu{\'e}, Nathalie},
title={Near-ultraviolet photon-counting dual-comb spectroscopy},
journal={Nature},
year={2024},
month={Mar},
day={01},
volume={627},
number={8003},
pages={289-294},
issn={1476-4687},
doi={10.1038/s41586-024-07094-9},
}

@article{McCauley:2024,
author = {John J. McCauley and Mark C. Phillips and Reagan R. D. Weeks and Yu Zhang and Sivanandan S. Harilal and R. Jason Jones},
journal = {Optica},
number = {4},
pages = {460--463},
publisher = {Optica Publishing Group},
title = {Dual-comb spectroscopy in the deep ultraviolet},
volume = {11},
month = {Apr},
year = {2024},
doi = {10.1364/OPTICA.516851},
}

@article{Chang:2024,
author = {Kristina F. Chang and Daniel M. B. Lesko and Carter Mashburn and Peter Chang and Eugene Tsao and Alexander J. Lind and Scott A. Diddams},
journal = {Opt. Lett.},
number = {7},
pages = {1684--1687},
publisher = {Optica Publishing Group},
title = {Multi-harmonic near-infrared-ultraviolet dual-comb spectrometer},
volume = {49},
month = {Apr},
year = {2024},
doi = {10.1364/OL.515776},
}

@Article{Wu:2024,
author={Wu, Tsung-Han
and Ledezma, Luis
and Fredrick, Connor
and Sekhar, Pooja
and Sekine, Ryoto
and Guo, Qiushi
and Briggs, Ryan M.
and Marandi, Alireza
and Diddams, Scott A.},
title={Visible-to-ultraviolet frequency comb generation in lithium niobate nanophotonic waveguides},
journal={Nature Photonics},
year={2024},
month={Mar},
day={01},
volume={18},
number={3},
pages={218-223},
issn={1749-4893},
doi={10.1038/s41566-023-01364-0},
}

@article{Schuster:2021,
    author = {V. Schuster and C. Liu and R. Klas and P. Dominguez and J. Rothhardt and J. Limpert and B. Bernhardt},
    journal = {Opt. Express},
    number = {14},
    pages = {21859--21875},
    title = {Ultraviolet dual comb spectroscopy: a roadmap},
    volume = {29},
    month = {Jul},
    year = {2021}
}

@Article{Galiter:2020,
    AUTHOR = {S. Galtier and P. Clément and P. Rairoux},
    TITLE = {Towards {DCS} in the {UV} Spectral Range for Remote Sensing of Atmospheric Trace Gases},
    JOURNAL = {Remote Sensing},
    VOLUME = {12},
    YEAR = {2020},
    NUMBER = {20},
    ARTICLE-NUMBER = {3444},
    doi = {10.3390/rs12203444}
}

@article{Coddington:2016,
author = {I. Coddington and N. Newbury and W. Swann},
journal = {Optica},
number = {4},
pages = {414--426},
publisher = {Optica Publishing Group},
title = {Dual-comb spectroscopy},
volume = {3},
year = {2016},
doi = {10.1364/OPTICA.3.000414}
}

@article{Schmidt:Espresso,
    author = {Schmidt, Tobias M and Reiners, Ansgar and Murphy, Michael T and Curto, Gaspare Lo and Martins, Carlos J A P and Huke, Philipp},
    title = {Validation of the ESPRESSO Wavelength Calibration Using Iodine Absorption Cell Spectra},
    journal = {Monthly Notices of the Royal Astronomical Society},
    pages = {staf588},
    year = {2025},
    month = {04},
    issn = {0035-8711},
    doi = {10.1093/mnras/staf588},
    url = {https://doi.org/10.1093/mnras/staf588},
    eprint = {https://academic.oup.com/mnras/advance-article-pdf/doi/10.1093/mnras/staf588/62941624/staf588.pdf},
}

@article{WalshGPU,
    author = {Walsh, Mathieu and Kasic, James and Cossel, Kevin and Genest, Jérôme},
    title = {Graphics card-based real-time processing for dual comb interferometry},
    journal = {Review of Scientific Instruments},
    volume = {95},
    number = {10},
    pages = {103005},
    year = {2024},
    month = {10},
    doi = {10.1063/5.0222548}
}

@article{Okubo_2015,
doi = {10.7567/APEX.8.082402},
url = {https://dx.doi.org/10.7567/APEX.8.082402},
year = {2015},
month = {jul},
publisher = {The Japan Society of Applied Physics},
volume = {8},
number = {8},
pages = {082402},
author = {Okubo, Sho and Iwakuni, Kana and Inaba, Hajime and Hosaka, Kazumoto and Onae, Atsushi and Sasada, Hiroyuki and Hong, Feng-Lei},
title = {Ultra-broadband dual-comb spectroscopy across 1.0–1.9 µm},
journal = {Applied Physics Express}
}

@article{SALAMI&ROSS,
title = {A molecular iodine atlas in ascii format},
journal = {Journal of Molecular Spectroscopy},
volume = {233},
number = {1},
pages = {157-159},
year = {2005},
issn = {0022-2852},
doi = {10.1016/j.jms.2005.06.002},
url = {https://www.sciencedirect.com/science/article/pii/S002228520500130X},
author = {Houssam Salami and Amanda J. Ross}
}

@article{Cole:Cepstral,
	title = {Baseline-free quantitative absorption spectroscopy based on cepstral analysis},
	volume = {27},
	copyright = {© 2019 Optical Society of America},
	issn = {1094-4087},
	url = {https://opg.optica.org/oe/abstract.cfm?uri=oe-27-26-37920},
	doi = {10.1364/OE.27.037920},
	language = {EN},
	number = {26},
	urldate = {2025-04-30},
	journal = {Optics Express},
	author = {Cole, Ryan K. and Makowiecki, Amanda S. and Hoghooghi, Nazanin and Rieker, Gregory B.},
	month = dec,
	year = {2019},
	note = {Publisher: Optica Publishing Group},
	pages = {37920--37939},
}

@ARTICLE{Vandaele2002,
  title     = "High‐resolution Fourier transform measurement of the {NO$_{2}$}
               visible and near‐infrared absorption cross sections: Temperature
               and pressure effects",
  author    = "Vandaele, A C and Hermans, C and Fally, S and Carleer, M and
               Colin, R and M{\'e}rienne, M-F and Jenouvrier, A and Coquart, B",
  journal   = "J. Geophys. Res.",
  publisher = "American Geophysical Union (AGU)",
  volume    =  107,
  number    = "D18",
  month     =  sep,
  year      =  2002,
  copyright = "http://onlinelibrary.wiley.com/termsAndConditions\#vor",
  language  = "en", 
  doi = {10.1029/2001JD000971}
}

@article{Ndengu2023,
  doi = {10.1021/acs.jpca.3c02832},
  url = {https://doi.org/10.1021/acs.jpca.3c02832},
  year = {2023},
  month = {june},
  publisher = {American Chemical Society ({ACS})},
  volume = {127},
  number = {29},
  pages = {6051--6062},
  author = {Steve Ndengu{\'{e}} and Ernesto Quintas-S{\'{a}}nchez and Richard Dawes and Christopher C. Blackstone and David L. Osborn},
  title = {Temperature Dependence of the Electronic Absorption Spectrum of {{NO}}2},
  journal = {The Journal of Physical Chemistry A}
}

@article{Ndengu2021,
  title = {The Low-Lying Electronic States of {{NO}}2: Potential Energy and Dipole Surfaces,  Bound States,  and Electronic Absorption Spectrum},
  volume = {125},
  ISSN = {1520-5215},
  url = {http://dx.doi.org/10.1021/acs.jpca.1c03482},
  DOI = {10.1021/acs.jpca.1c03482},
  number = {25},
  journal = {The Journal of Physical Chemistry A},
  publisher = {American Chemical Society (ACS)},
  author = {Ndengué,  Steve and Quintas-Sánchez,  Ernesto and Dawes,  Richard and Osborn,  David},
  year = {2021},
  month = jun,
  pages = {5519–5533}
}

@article{Schmidt2021,
  title = {Fundamental physics with ESPRESSO: Towards an accurate wavelength calibration for a precision test of the fine-structure constant},
  volume = {646},
  ISSN = {1432-0746},
  url = {http://dx.doi.org/10.1051/0004-6361/202039345},
  DOI = {10.1051/0004-6361/202039345},
  journal = {Astronomy \& Astrophysics},
  publisher = {EDP Sciences},
  author = {Schmidt,  Tobias M. and Molaro,  Paolo and Murphy,  Michael T. and Lovis,  Christophe and Cupani,  Guido and Cristiani,  Stefano and Pepe,  Francesco A. and Rebolo,  Rafael and Santos,  Nuno C. and Abreu,  Manuel and Adibekyan,  Vardan and Alibert,  Yann and Aliverti,  Matteo and Allart,  Romain and Allende Prieto,  Carlos and Alves,  David and Baldini,  Veronica and Broeg,  Christopher and Cabral,  Alexandre and Calderone,  Giorgio and Cirami,  Roberto and Coelho,  João and Coretti,  Igor and D’Odorico,  Valentina and Di Marcantonio,  Paolo and Ehrenreich,  David and Figueira,  Pedro and Genoni,  Matteo and Génova Santos,  Ricardo and González Hernández,  Jonay I. and Kerber,  Florian and Landoni,  Marco and Leite,  Ana C. O. and Lizon,  Jean-Louis and Lo Curto,  Gaspare and Manescau,  Antonio and Martins,  Carlos J. A. P. and Megévand,  Denis and Mehner,  Andrea and Micela,  Giuseppina and Modigliani,  Andrea and Monteiro,  Manuel and Monteiro,  Mario J. P. F. G. and Mueller,  Eric and Nunes,  Nelson J. and Oggioni,  Luca and Oliveira,  António and Pariani,  Giorgio and Pasquini,  Luca and Redaelli,  Edoardo and Riva,  Marco and Santos,  Pedro and Sosnowska,  Danuta and Sousa,  Sérgio G. and Sozzetti,  Alessandro and Suárez Mascareño,  Alejandro and Udry,  Stéphane and Zapatero Osorio,  Maria-Rosa and Zerbi,  Filippo},
  year = {2021},
  month = feb,
  pages = {A144}
}

@article{Timmers:18,
author = {Henry Timmers and Abijith Kowligy and Alex Lind and Flavio C. Cruz and Nima Nader and Myles Silfies and Gabriel Ycas and Thomas K. Allison and Peter G. Schunemann and Scott B. Papp and Scott A. Diddams},
journal = {Optica},
number = {6},
pages = {727--732},
publisher = {Optica Publishing Group},
title = {Molecular fingerprinting with bright, broadband infrared frequency combs},
volume = {5},
month = {Jun},
year = {2018},
url = {https://opg.optica.org/optica/abstract.cfm?URI=optica-5-6-727},
doi = {10.1364/OPTICA.5.000727},
}

@article{Bera2008,
  title = {Born{{-O}}ppenheimer Symmetry Breaking in the {{C}}-State of {{NO}}2: Importance of Static and Dynamic Correlation Effects},
  volume = {112},
  ISSN = {1520-5215},
  url = {http://dx.doi.org/10.1021/jp077561y},
  DOI = {10.1021/jp077561y},
  number = {12},
  journal = {The Journal of Physical Chemistry A},
  publisher = {American Chemical Society (ACS)},
  author = {Bera,  Partha P. and Yamaguchi,  Yukio and Schaefer,  Henry F. and Crawford,  T. Daniel},
  year = {2008},
  month = mar,
  pages = {2669–2676}
}

@article{RevModPhys.87.637,
  title = {Optical atomic clocks},
  author = {Ludlow, Andrew D. and Boyd, Martin M. and Ye, Jun and Peik, E. and Schmidt, P. O.},
  journal = {Rev. Mod. Phys.},
  volume = {87},
  issue = {2},
  pages = {637--701},
  numpages = {65},
  year = {2015},
  month = {Jun},
  publisher = {American Physical Society},
  doi = {10.1103/RevModPhys.87.637},
  url = {https://link.aps.org/doi/10.1103/RevModPhys.87.637}
}

@article{Cunge_2011,
doi = {10.1088/0022-3727/44/12/122001},
url = {https://dx.doi.org/10.1088/0022-3727/44/12/122001},
year = {2011},
month = {mar},
publisher = {},
volume = {44},
number = {12},
pages = {122001},
author = {Cunge, G and Fouchier, M and Brihoum, M and Bodart, P and Touzeau, M and Sadeghi, N},
title = {Vacuum {{UV}} broad-band absorption spectroscopy: a powerful diagnostic tool for reactive plasma monitoring},
journal = {Journal of Physics D: Applied Physics}
}

@article{Adamovich_2022,
doi = {10.1088/1361-6463/ac5e1c},
url = {https://dx.doi.org/10.1088/1361-6463/ac5e1c},
year = {2022},
month = {jul},
publisher = {IOP Publishing},
volume = {55},
number = {37},
pages = {373001},
author = {Adamovich, I and Agarwal, S and Ahedo, E and Alves, L L and Baalrud, S and Babaeva, N and Bogaerts, A and Bourdon, A and Bruggeman, P J and Canal, C and Choi, E H and Coulombe, S and Donkó, Z and Graves, D B and Hamaguchi, S and Hegemann, D and Hori, M and Kim, H-H and Kroesen, G M W and Kushner, M J and Laricchiuta, A and Li, X and Magin, T E and Mededovic Thagard, S and Miller, V and Murphy, A B and Oehrlein, G S and Puac, N and Sankaran, R M and Samukawa, S and Shiratani, M and Šimek, M and Tarasenko, N and Terashima, K and Thomas Jr, E and Trieschmann, J and Tsikata, S and Turner, M M and van der Walt, I J and van de Sanden, M C M and von Woedtke, T},
title = {The 2022 Plasma Roadmap: low temperature plasma science and technology},
journal = {Journal of Physics D: Applied Physics}
}

@article{Stroud2025,
  title = {Frequency-doubled chirped-pulse dual-comb generation in the near-{{UV}}: combined vs separated beam investigations of {{R}}b atoms and {{NO}}2 near 420 nm},
  volume = {15},
  ISSN = {2045-2322},
  url = {http://dx.doi.org/10.1038/s41598-025-00684-1},
  DOI = {10.1038/s41598-025-00684-1},
  number = {1},
  journal = {Scientific Reports},
  publisher = {Springer Science and Business Media LLC},
  author = {Stroud,  Jasper R. and Plusquellic,  David F.},
  year = {2025},
  month = may 
}

@article{Hofer2025,
  title = {Free-Running Deep-{{UV}} Dual-Comb Spectroscopy},
  ISSN = {1361-6455},
  url = {http://dx.doi.org/10.1088/1361-6455/add4d7},
  DOI = {10.1088/1361-6455/add4d7},
  journal = {Journal of Physics B: Atomic,  Molecular and Optical Physics},
  publisher = {IOP Publishing},
  author = {Hofer,  Tobias and Hehl,  Gregor Fabian Magnus and Meyer,  Johann Gabriel and Pronin,  Oleg},
  year = {2025},
  month = may 
}

@article{
Li2025,
author = {Quanming Li  and Hanze Bai  and Xiaodan Teng  and Hongshan Chen  and Haijing Mai  and Zhitao Zhang  and Jinwei Zhang  and Hongwen Xuan },
title = {Deep Ultraviolet Dual Comb from a Thin-Disk Laser},
journal = {Ultrafast Science},
volume = {5},
number = {},
pages = {0087},
year = {2025},
doi = {10.34133/ultrafastscience.0087},
URL = {https://spj.science.org/doi/abs/10.34133/ultrafastscience.0087},
eprint = {https://spj.science.org/doi/pdf/10.34133/ultrafastscience.0087}}

@misc{fuerst2025,
      title={Ultra-resolution photochemical sensing}, 
      author={Lukas Fuerst and Alexander Eber and Mithun Pal and Emily Hruska and Clemens Hofmann and Iouli Gordon and Martin Schultze and Rolf Breinbauer and Birgitta Bernhardt},
      year={2025},
      eprint={2501.07350},
      archivePrefix={arXiv},
      primaryClass={physics.optics},
      url={https://arxiv.org/abs/2501.07350}, 
      doi = {10.48550/arXiv.2501.07350}
}

@article{Muraviev2024,
  title = {Dual-frequency-comb {{UV}} spectroscopy with one million resolved comb lines},
  volume = {11},
  ISSN = {2334-2536},
  url = {http://dx.doi.org/10.1364/OPTICA.536971},
  DOI = {10.1364/optica.536971},
  number = {11},
  journal = {Optica},
  publisher = {Optica Publishing Group},
  author = {Muraviev,  Andrey and Konnov,  Dmitrii and Vasilyev,  Sergey and Vodopyanov,  Konstantin L.},
  year = {2024},
  month = oct,
  pages = {1486}
}

@article{Ideguchi2012,
  title = {Adaptive dual-comb spectroscopy in the green region},
  volume = {37},
  ISSN = {1539-4794},
  url = {http://dx.doi.org/10.1364/OL.37.004847},
  DOI = {10.1364/ol.37.004847},
  number = {23},
  journal = {Optics Letters},
  publisher = {Optica Publishing Group},
  author = {Ideguchi,  T. and Poisson,  A. and Guelachvili,  G. and H\"{a}nsch,  T. W. and Picqué,  N.},
  year = {2012},
  month = nov,
  pages = {4847}
}

@article{Tian2024,
  title = {Broadband,  high-power optical frequency combs covering visible to near-infrared spectral range},
  volume = {49},
  ISSN = {1539-4794},
  url = {http://dx.doi.org/10.1364/OL.514182},
  DOI = {10.1364/ol.514182},
  number = {3},
  journal = {Optics Letters},
  publisher = {Optica Publishing Group},
  author = {Tian,  Haochen and Zhu,  Ruichen and Li,  Runmin and Xing,  Sida and Schibli,  Thomas R. and Minoshima,  Kaoru},
  year = {2024},
  month = jan,
  pages = {538}
}

@article{
Pal2025,
author = {Mithun Pal  and Alexander Eber  and Lukas Fürst  and Emily Hruska  and Marcus Ossiander  and Birgitta Bernhardt },
title = {Phase-Locked Feed-Forward Stabilization for Dual-Comb Spectroscopy},
journal = {Ultrafast Science},
volume = {5},
number = {},
pages = {0098},
year = {2025},
doi = {10.34133/ultrafastscience.0098},
URL = {https://spj.science.org/doi/abs/10.34133/ultrafastscience.0098},
eprint = {https://spj.science.org/doi/pdf/10.34133/ultrafastscience.0098}}

@article{Lind2020,
  title = {Mid-Infrared Frequency Comb Generation and Spectroscopy with Few-Cycle Pulses and ${\ensuremath{\chi}}^{(2)}$ Nonlinear Optics},
  author = {Lind, Alexander J. and Kowligy, Abijith and Timmers, Henry and Cruz, Flavio C. and Nader, Nima and Silfies, Myles C. and Allison, Thomas K. and Diddams, Scott A.},
  journal = {Phys. Rev. Lett.},
  volume = {124},
  issue = {13},
  pages = {133904},
  numpages = {6},
  year = {2020},
  month = {Apr},
  publisher = {American Physical Society},
  doi = {10.1103/PhysRevLett.124.133904},
  url = {https://link.aps.org/doi/10.1103/PhysRevLett.124.133904}
}

@article{Hoghooghi:24,
author = {Nazanin Hoghooghi and Peter Chang and Scott Egbert and Matt Burch and Rizwan Shaik and Scott A. Diddams and Patrick Lynch and Gregory B. Rieker},
journal = {Optica},
number = {6},
pages = {876--882},
publisher = {Optica Publishing Group},
title = {{{GH}}z repetition rate mid-infrared frequency comb spectroscopy of fast chemical reactions},
volume = {11},
month = {Jun},
year = {2024},
url = {https://opg.optica.org/optica/abstract.cfm?URI=optica-11-6-876},
doi = {10.1364/OPTICA.521655},
}

@article{
Herman2021,
author = {Daniel I. Herman  and Chinthaka Weerasekara  and Lindsay C. Hutcherson  and Fabrizio R. Giorgetta  and Kevin C. Cossel  and Eleanor M. Waxman  and Gabriel M. Colacion  and Nathan R. Newbury  and Stephen M. Welch  and Brett D. DePaola  and Ian Coddington  and Eduardo A. Santos  and Brian R. Washburn },
title = {Precise multispecies agricultural gas flux determined using broadband open-path dual-comb spectroscopy},
journal = {Science Advances},
volume = {7},
number = {14},
pages = {eabe9765},
year = {2021},
doi = {10.1126/sciadv.abe9765},
URL = {https://www.science.org/doi/abs/10.1126/sciadv.abe9765},
eprint = {https://www.science.org/doi/pdf/10.1126/sciadv.abe9765}}

@article{
Huan:CleanAir,
author = {Wei Huang  and Hongbing Xu  and Jing Wu  and Minghui Ren  and Yang Ke  and Jie Qiao },
title = {Toward cleaner air and better health: Current state, challenges, and priorities},
journal = {Science},
volume = {385},
number = {6707},
pages = {386-390},
doi = {10.1126/science.adp7832},
year = {2024}
}

@article{
Lobell:CropGrowth,
author = {David B. Lobell  and Stefania Di Tommaso  and Jennifer A. Burney },
title = {Globally ubiquitous negative effects of nitrogen dioxide on crop growth},
journal = {Science Advances},
volume = {8},
number = {22},
pages = {eabm9909},
year = {2022},
doi = {10.1126/sciadv.abm9909}
}

@article{Newbury:sensitivity,
author = {Nathan R. Newbury and Ian Coddington and William Swann},
journal = {Opt. Express},
number = {8},
pages = {7929--7945},
publisher = {Optica Publishing Group},
title = {Sensitivity of coherent dual-comb spectroscopy},
volume = {18},
month = {Apr},
year = {2010},
url = {https://opg.optica.org/oe/abstract.cfm?URI=oe-18-8-7929},
doi = {10.1364/OE.18.007929},
}

@article{Yun:23,
author = {David Yun and Walter B. Sabin and Sean C. Coburn and Nazanin Hoghooghi and Jacob J. France and Mark A. Hagenmaier and Kristin M. Rice and Jeffrey M. Donbar and Gregory B. Rieker},
journal = {Opt. Express},
number = {25},
pages = {42571--42580},
publisher = {Optica Publishing Group},
title = {Thermometry and velocimetry in a ramjet using dual comb spectroscopy of the {{O}}2 {{A}}-band},
volume = {31},
month = {Dec},
year = {2023},
url = {https://opg.optica.org/oe/abstract.cfm?URI=oe-31-25-42571},
doi = {10.1364/OE.507647},
}

@article{Roslund2024,
  title = {Optical clocks at sea},
  volume = {628},
  ISSN = {1476-4687},
  url = {http://dx.doi.org/10.1038/s41586-024-07225-2},
  DOI = {10.1038/s41586-024-07225-2},
  number = {8009},
  journal = {Nature},
  publisher = {Springer Science and Business Media LLC},
  author = {Roslund,  Jonathan D. and Cing\"{o}z,  Arman and Lunden,  William D. and Partridge,  Guthrie B. and Kowligy,  Abijith S. and Roller,  Frank and Sheredy,  Daniel B. and Skulason,  Gunnar E. and Song,  Joe P. and Abo-Shaeer,  Jamil R. and Boyd,  Martin M.},
  year = {2024},
  month = apr,
  pages = {736–740}
}

@article{MenloSoundingRocket,
  title = {Iodine Frequency Reference on a Sounding Rocket},
  author = {D\"oringshoff, Klaus and Gutsch, Franz B. and Schkolnik, Vladimir and K\"urbis, Christian and Oswald, Markus and Pr\"obster, Benjamin and Kovalchuk, Evgeny V. and Bawamia, Ahmad and Smol, Robert and Schuldt, Thilo and Lezius, Matthias and Holzwarth, Ronald and Wicht, Andreas and Braxmaier, Claus and Krutzik, Markus and Peters, Achim},
  journal = {Phys. Rev. Appl.},
  volume = {11},
  issue = {5},
  pages = {054068},
  numpages = {9},
  year = {2019},
  month = {May},
  publisher = {American Physical Society},
  doi = {10.1103/PhysRevApplied.11.054068},
  url = {https://link.aps.org/doi/10.1103/PhysRevApplied.11.054068}
}

@article{FERNANDEZ2023,
title = {High resolution laser spectroscopy of iodine molecule in the 14400–14600 cm-1 range},
journal = {Journal of Molecular Spectroscopy},
volume = {395},
pages = {111789},
year = {2023},
issn = {0022-2852},
doi = {10.1016/j.jms.2023.111789},
url = {https://www.sciencedirect.com/science/article/pii/S0022285223000541},
author = {David Rodríguez Fernández and Manuel Alejandro Lefrán Torres and Marcos Roberto Cardoso and Jorge Douglas Massayuki Kondo and Luis Gustavo Marcassa}
}

@article{Torres2022,
title = {High resolution laser spectroscopy of iodine molecule in the 14600–14710cm-1 range},
journal = {Journal of Molecular Spectroscopy},
volume = {387},
pages = {111668},
year = {2022},
issn = {0022-2852},
doi = {10.1016/j.jms.2022.111668},
url = {https://www.sciencedirect.com/science/article/pii/S0022285222000868},
author = {Manuel Alejandro {Lefrán Torres} and David Rodríguez Fernández and Marcos Roberto Cardoso and Luis Gustavo Marcassa}
}

@article{Reiners2024,
	author = {{Reiners, A.} and {Debus, M.} and {Schäfer, S.} and {Tiemann, E.} and {Zechmeister, M.}},
	title = {Accurate calibration spectra for precision radial velocities - Iodine absorption referenced by a laser frequency comb},
	DOI= "10.1051/0004-6361/202451389",
	url= "https://doi.org/10.1051/0004-6361/202451389",
	journal = {A \& A},
	year = 2024,
	volume = 690,
	pages = "A210",
}

@article{Sinclair:14,
author = {L. C. Sinclair and I. Coddington and W. C. Swann and G. B. Rieker and A. Hati and K. Iwakuni and N. R. Newbury},
journal = {Opt. Express},
number = {6},
pages = {6996--7006},
publisher = {Optica Publishing Group},
title = {Operation of an optically coherent frequency comb outside the metrology lab},
volume = {22},
month = {Mar},
year = {2014},
url = {https://opg.optica.org/oe/abstract.cfm?URI=oe-22-6-6996},
doi = {10.1364/OE.22.006996},
}

@article{Chen2024,
  title = {High-performance and fabrication-tolerant edge coupler on thin film lithium niobate based on a three-dimensional inverse taper},
  volume = {9},
  ISSN = {2378-0967},
  url = {http://dx.doi.org/10.1063/5.0224269},
  DOI = {10.1063/5.0224269},
  number = {11},
  journal = {APL Photonics},
  publisher = {AIP Publishing},
  author = {Chen,  Bin and Ruan,  Ziliang and Wang,  Mai and Gong,  Shengqi and Liu,  Liu},
  year = {2024},
  month = nov 
}

@article{Fan:24,
author = {Weichen Fan and Markus Ludwig and Ian Rousseau and Ivo Arabadzhiev and Bastian Ruhnke and Thibault Wildi and Tobias Herr},
journal = {Optica},
number = {8},
pages = {1175--1181},
publisher = {Optica Publishing Group},
title = {Supercontinua from integrated gallium nitride waveguides},
volume = {11},
month = {Aug},
year = {2024},
url = {https://opg.optica.org/optica/abstract.cfm?URI=optica-11-8-1175},
doi = {10.1364/OPTICA.528341},
}

@article{Guay:18,
author = {Philippe Guay and J\'{e}r\^{o}me Genest and Adam J. Fleisher},
journal = {Opt. Lett.},
number = {6},
pages = {1407--1410},
publisher = {Optica Publishing Group},
title = {Precision spectroscopy of {{H13CN}} using a free-running, all-fiber dual electro-optic frequency comb system},
volume = {43},
month = {Mar},
year = {2018},
url = {https://opg.optica.org/ol/abstract.cfm?URI=ol-43-6-1407},
doi = {10.1364/OL.43.001407},
}

@article{Guay:19,
author = {Philippe Guay and Nicolas Bourbeau H\'{e}bert and Vincent Michaud-Belleau and David G. Lancaster and J\'{e}r\^{o}me Genest},
journal = {Opt. Lett.},
number = {17},
pages = {4375--4378},
publisher = {Optica Publishing Group},
title = {Methane spectroscopy using a free-running chip-based dual-comb laser},
volume = {44},
month = {Sep},
year = {2019},
url = {https://opg.optica.org/ol/abstract.cfm?URI=ol-44-17-4375},
doi = {10.1364/OL.44.004375},
}

@article{Muraviev2020,
  title = {Broadband high-resolution molecular spectroscopy with interleaved mid-infrared frequency combs},
  volume = {10},
  ISSN = {2045-2322},
  url = {http://dx.doi.org/10.1038/s41598-020-75704-3},
  DOI = {10.1038/s41598-020-75704-3},
  number = {1},
  journal = {Scientific Reports},
  publisher = {Springer Science and Business Media LLC},
  author = {Muraviev,  A. V. and Konnov,  D. and Vodopyanov,  K. L.},
  year = {2020},
  month = oct 
}

@article{Weeks2022,
  title = {Multi-species temperature and number density analysis of a laser-produced plasma using dual-comb spectroscopy},
  volume = {131},
  ISSN = {1089-7550},
  url = {http://dx.doi.org/10.1063/5.0094213},
  DOI = {10.1063/5.0094213},
  number = {22},
  journal = {Journal of Applied Physics},
  publisher = {AIP Publishing},
  author = {Weeks,  Reagan R. D. and Zhang,  Yu and Harilal,  Sivanandan S. and Phillips,  Mark C. and Jones,  R. Jason},
  year = {2022},
  month = jun 
}

@article{Lesko2022,
  title = {High-sensitivity frequency comb carrier-envelope-phase metrology in solid state high harmonic generation},
  volume = {9},
  ISSN = {2334-2536},
  url = {http://dx.doi.org/10.1364/OPTICA.465709},
  DOI = {10.1364/optica.465709},
  number = {10},
  journal = {Optica},
  publisher = {Optica Publishing Group},
  author = {Lesko,  Daniel M. B. and Chang,  Kristina F. and Diddams,  Scott A.},
  year = {2022},
  month = oct,
  pages = {1156}
}

@inproceedings{Osterman2007,
  title = {A proposed laser frequency comb-based wavelength reference for high-resolution spectroscopy},
  ISSN = {0277-786X},
  url = {http://dx.doi.org/10.1117/12.734193},
  DOI = {10.1117/12.734193},
  booktitle = {Techniques and Instrumentation for Detection of Exoplanets III},
  publisher = {SPIE},
  author = {Osterman,  Steve and Diddams,  Scott and Beasley,  Matthew and Froning,  Cynthia and Hollberg,  Leo and MacQueen,  Phillip and Mbele,  Vela and Weiner,  Andrew},
  editor = {Coulter,  Daniel R.},
  year = {2007},
  month = sep 
}

@article{Wilken2012,
  title = {A spectrograph for exoplanet observations calibrated at the centimetre-per-second level},
  volume = {485},
  ISSN = {1476-4687},
  url = {http://dx.doi.org/10.1038/nature11092},
  DOI = {10.1038/nature11092},
  number = {7400},
  journal = {Nature},
  publisher = {Springer Science and Business Media LLC},
  author = {Wilken,  Tobias and Curto,  Gaspare Lo and Probst,  Rafael A. and Steinmetz,  Tilo and Manescau,  Antonio and Pasquini,  Luca and González Hernández,  Jonay I. and Rebolo,  Rafael and H\"{a}nsch,  Theodor W. and Udem,  Thomas and Holzwarth,  Ronald},
  year = {2012},
  month = may,
  pages = {611–614}
}

@article{Glenday2015,
  title = {Operation of a broadband visible-wavelength astro-comb with a high-resolution astrophysical spectrograph},
  volume = {2},
  ISSN = {2334-2536},
  url = {http://dx.doi.org/10.1364/OPTICA.2.000250},
  DOI = {10.1364/optica.2.000250},
  number = {3},
  journal = {Optica},
  publisher = {Optica Publishing Group},
  author = {Glenday,  Alexander G. and Li,  Chih-Hao and Langellier,  Nicholas and Chang,  Guoqing and Chen,  Li-Jin and Furesz,  Gabor and Zibrov,  Alexander A. and K\"{a}rtner,  Franz and Phillips,  David F. and Sasselov,  Dimitar and Szentgyorgyi,  Andrew and Walsworth,  Ronald L.},
  year = {2015},
  month = mar,
  pages = {250}
}

@article{Genier2021,
  title = {Ultra-flat,  low-noise,  and linearly polarized fiber supercontinuum source covering 670–1390 nm},
  volume = {46},
  ISSN = {1539-4794},
  url = {http://dx.doi.org/10.1364/OL.420676},
  DOI = {10.1364/ol.420676},
  number = {8},
  journal = {Optics Letters},
  publisher = {Optica Publishing Group},
  author = {Genier,  Etienne and Grelet,  Sacha and Engelsholm,  Rasmus D. and Bowen,  Patrick and Moselund,  Peter M. and Bang,  Ole and Dudley,  John M. and Sylvestre,  Thibaut},
  year = {2021},
  month = apr,
  pages = {1820}
}

@article{Carlson2018,
  title = {Ultrafast electro-optic light with subcycle control},
  volume = {361},
  ISSN = {1095-9203},
  url = {http://dx.doi.org/10.1126/science.aat6451},
  DOI = {10.1126/science.aat6451},
  number = {6409},
  journal = {Science},
  publisher = {American Association for the Advancement of Science (AAAS)},
  author = {Carlson,  David R. and Hickstein,  Daniel D. and Zhang,  Wei and Metcalf,  Andrew J. and Quinlan,  Franklyn and Diddams,  Scott A. and Papp,  Scott B.},
  year = {2018},
  month = sep,
  pages = {1358–1363}
}

@article{Huber2000,
  title = {Generation and field-resolved detection of femtosecond electromagnetic pulses tunable up to 41 {{TH}}z},
  volume = {76},
  ISSN = {1077-3118},
  url = {http://dx.doi.org/10.1063/1.126625},
  DOI = {10.1063/1.126625},
  number = {22},
  journal = {Applied Physics Letters},
  publisher = {AIP Publishing},
  author = {Huber,  R. and Brodschelm,  A. and Tauser,  F. and Leitenstorfer,  A.},
  year = {2000},
  month = may,
  pages = {3191–3193}
}

@article{Kowalczyk2023,
  title = {Ultra-CEP-stable single-cycle pulses at 2.2µm},
  volume = {10},
  ISSN = {2334-2536},
  url = {http://dx.doi.org/10.1364/OPTICA.481673},
  DOI = {10.1364/optica.481673},
  number = {6},
  journal = {Optica},
  publisher = {Optica Publishing Group},
  author = {Kowalczyk,  Maciej and Nagl,  Nathalie and Steinleitner,  Philipp and Karpowicz,  Nicholas and Pervak,  Vladimir and Głuszek,  Aleksander and Hudzikowski,  Arkadiusz and Krausz,  Ferenc and Mak,  Ka Fai and Weigel,  Alexander},
  year = {2023},
  month = jun,
  pages = {801}
}

@article{Abbas2019,
  title = {Time-resolved mid-infrared dual-comb spectroscopy},
  volume = {9},
  ISSN = {2045-2322},
  url = {http://dx.doi.org/10.1038/s41598-019-53825-8},
  DOI = {10.1038/s41598-019-53825-8},
  number = {1},
  journal = {Scientific Reports},
  publisher = {Springer Science and Business Media LLC},
  author = {Abbas,  Muhammad A. and Pan,  Qing and Mandon,  Julien and Cristescu,  Simona M. and Harren,  Frans J. M. and Khodabakhsh,  Amir},
  year = {2019},
  month = nov 
}

@article{Yu2018,
  title = {Silicon-chip-based mid-infrared dual-comb spectroscopy},
  volume = {9},
  ISSN = {2041-1723},
  url = {http://dx.doi.org/10.1038/s41467-018-04350-1},
  DOI = {10.1038/s41467-018-04350-1},
  number = {1},
  journal = {Nature Communications},
  publisher = {Springer Science and Business Media LLC},
  author = {Yu,  Mengjie and Okawachi,  Yoshitomo and Griffith,  Austin G. and Picqué,  Nathalie and Lipson,  Michal and Gaeta,  Alexander L.},
  year = {2018},
  month = may 
}

@article{BOGUMIL2003,
title = {Measurements of molecular absorption spectra with the SCIAMACHY pre-flight model: instrument characterization and reference data for atmospheric remote-sensing in the 230–2380 nm region},
journal = {Journal of Photochemistry and Photobiology A: Chemistry},
volume = {157},
number = {2},
pages = {167-184},
year = {2003},
issn = {1010-6030},
doi = {10.1016/S1010-6030(03)00062-5},
author = {K Bogumil and J Orphal and T Homann and S Voigt and P Spietz and O.C Fleischmann and A Vogel and M Hartmann and H Kromminga and H Bovensmann and J Frerick and J.P Burrows}
}

@article{Nizkorodov2004,
  title = {Temperature and Pressure Dependence of High-Resolution Air-Broadened Absorption Cross Sections of {NO}2 (415-525 nm)},
  volume = {108},
  issn = {1520-5215},
  url = {http://dx.doi.org/10.1021/jp049461n},
  doi = {10.1021/jp049461n},
  number = {22},
  journal = {The Journal of Physical Chemistry A},
  publisher = {American Chemical Society (ACS)},
  author = {Nizkorodov,  S. A. and Sander,  S. P. and Brown,  L. R.},
  year = {2004},
  month = {may},
  pages = {4864–4872}
}

@article{Reinhardt2007,
  title = {Absolute frequency measurements and comparisons in iodine at 735nm and 772nm},
  volume = {274},
  issn = {0030-4018},
  url = {http://dx.doi.org/10.1016/j.optcom.2007.02.050},
  doi = {10.1016/j.optcom.2007.02.050},
  number = {2},
  journal = {Optics Communications},
  publisher = {Elsevier BV},
  author = {Reinhardt,  S. and Bernhardt,  B. and Geppert,  C. and Holzwarth,  R. and Huber,  G. and Karpuk,  S. and Miski-Oglu,  N. and N\"{o}rtersh\"{a}user,  W. and Novotny,  C. and Udem,  Th.},
  year = {2007},
  month = {jun},
  pages = {354–360}
}

@article{Hebert:18,
author = {Nicolas Bourbeau H\'{e}bert and David G. Lancaster and Vincent Michaud-Belleau and George Y. Chen and J\'{e}r\^{o}me Genest},
journal = {Opt. Lett.},
number = {8},
pages = {1814--1817},
publisher = {Optica Publishing Group},
title = {Highly coherent free-running dual-comb chip platform},
volume = {43},
month = {Apr},
year = {2018},
url = {https://opg.optica.org/ol/abstract.cfm?URI=ol-43-8-1814},
doi = {10.1364/OL.43.001814},
}

@article{Hebert:17,
author = {Nicolas Bourbeau H\'{e}bert and J\'{e}r\^{o}me Genest and Jean-Daniel Desch\^{e}nes and Hugo Bergeron and George Y. Chen and Champak Khurmi and David G. Lancaster},
journal = {Opt. Express},
number = {7},
pages = {8168--8179},
publisher = {Optica Publishing Group},
title = {Self-corrected chip-based dual-comb spectrometer},
volume = {25},
month = {Apr},
year = {2017},
url = {https://opg.optica.org/oe/abstract.cfm?URI=oe-25-7-8168},
doi = {10.1364/OE.25.008168},
}

@article{Roy:12,
author = {Julien Roy and Jean-Daniel Desch\^{e}nes and Simon Potvin and J\'{e}r\^{o}me Genest},
journal = {Opt. Express},
number = {20},
pages = {21932--21939},
publisher = {Optica Publishing Group},
title = {Continuous real-time correction and averaging for frequency comb interferometry},
volume = {20},
month = {Sep},
year = {2012},
url = {https://opg.optica.org/oe/abstract.cfm?URI=oe-20-20-21932},
doi = {10.1364/OE.20.021932},
}

@article{Ycas2018,
  title = {High-coherence mid-infrared dual-comb spectroscopy spanning 2.6 to 5.2 $\mu$m},
  volume = {12},
  ISSN = {1749-4893},
  url = {http://dx.doi.org/10.1038/s41566-018-0114-7},
  DOI = {10.1038/s41566-018-0114-7},
  number = {4},
  journal = {Nature Photonics},
  publisher = {Springer Science and Business Media LLC},
  author = {Ycas,  Gabriel and Giorgetta,  Fabrizio R. and Baumann,  Esther and Coddington,  Ian and Herman,  Daniel and Diddams,  Scott A. and Newbury,  Nathan R.},
  year = {2018},
  month = mar,
  pages = {202–208}
}

@article{steckSodium,
  title={Sodium {{D}} line data},
  author={Steck, Daniel A},
  journal={Report, Los Alamos National Laboratory, Los Alamos},
  volume={124},
  pages={74},
  year={2000},
  url={https://steck.us/alkalidata/}
}

@article{steckRubidium87,
  title={Rubidium 87 {{D}} line data},
  author={Steck, Daniel A},
  year={2001},  
  journal={Report, Los Alamos National Laboratory, Los Alamos},
  url={https://steck.us/alkalidata/}
}

@article{steckRubidium85,
  title={Rubidium 85 {{D}} line data},
  author={Steck, Daniel A},
  year={2008},  
  journal={Report, Los Alamos National Laboratory, Los Alamos},
  url={https://steck.us/alkalidata/}
}

@article{FanPRL,
  title = {Spectral Dynamics in Broadband Frequency Combs with Overlapping Harmonics},
  author = {Fan, Weichen and Ayhan, Furkan and Wildi, Thibault and Volkov, Mikhail and Seer, Ali and Ludwig, Markus and Voumard, Thibault and Brodschelm, Andreas and Brasch, Victor and Villanueva, Luis Guillermo and Herr, Tobias},
  journal = {Phys. Rev. Lett.},
  volume = {135},
  issue = {21},
  pages = {213801},
  numpages = {6},
  year = {2025},
  month = {Nov},
  publisher = {American Physical Society},
  doi = {10.1103/kgd7-52hy},
  url = {https://link.aps.org/doi/10.1103/kgd7-52hy}
}

@article{juncar1981,
  title={Absolute determination of the wavelengths of the sodium {{D}}1 and {{D}}2 lines by using a cw tunable dye laser stabilized on iodine},
  author={Juncar, P and Pinard, J and Hamon, J and Chartier, A},
  journal={Metrologia},
  volume={17},
  number={3},
  pages={77},
  year={1981},
  publisher={IOP Publishing},
  doi = {10.1088/0026-1394/17/3/001}
}

@article{Eber:25,
author = {Alexander Eber and Christoph Gruber and Martin Schultze and Birgitta Bernhardt and Marcus Ossiander},
journal = {Opt. Express},
number = {17},
pages = {35314--35325},
publisher = {Optica Publishing Group},
title = {Streaming self-corrected dual-comb spectrometer},
volume = {33},
month = {Aug},
year = {2025},
url = {https://opg.optica.org/oe/abstract.cfm?URI=oe-33-17-35314},
doi = {10.1364/OE.569404},
}

@article{KONNOV2025,
title = {High-resolution (8–16 {{MH}}z) rovibrational absorption spectra of low-pressure methanol, ethanol, isoprene, and dimethyl sulfide at 700–1500 cm-1 measured via dual-comb spectroscopy},
journal = {Journal of Quantitative Spectroscopy and Radiative Transfer},
volume = {347},
pages = {109690},
year = {2025},
issn = {0022-4073},
doi = {10.1016/j.jqsrt.2025.109690},
url = {https://www.sciencedirect.com/science/article/pii/S0022407325003528},
author = {D. Konnov and A. Muraviev and K.L. Vodopyanov}
}

@article{LindPRL,
  title = {Mid-Infrared Frequency Comb Generation and Spectroscopy with Few-Cycle Pulses and ${\ensuremath{\chi}}^{(2)}$ Nonlinear Optics},
  author = {Lind, Alexander J. and Kowligy, Abijith and Timmers, Henry and Cruz, Flavio C. and Nader, Nima and Silfies, Myles C. and Allison, Thomas K. and Diddams, Scott A.},
  journal = {Phys. Rev. Lett.},
  volume = {124},
  issue = {13},
  pages = {133904},
  numpages = {6},
  year = {2020},
  month = {Apr},
  publisher = {American Physical Society},
  doi = {10.1103/PhysRevLett.124.133904},
  url = {https://link.aps.org/doi/10.1103/PhysRevLett.124.133904}
}

@article{Kirchner:25,
author = {Adrian Kirchner and Alexander Eber and Lukas F\"{u}rst and Emily Hruska and Michael H. Frosz and Francesco Tani and Birgitta Bernhardt},
journal = {Opt. Express},
number = {4},
pages = {7005--7015},
publisher = {Optica Publishing Group},
title = {Ultra-broadband {{UV/VIS}} spectroscopy enabled by resonant dispersive wave emission of a frequency comb},
volume = {33},
month = {Feb},
year = {2025},
url = {https://opg.optica.org/oe/abstract.cfm?URI=oe-33-4-7005},
doi = {10.1364/OE.546751},
}

@article{Siddons2008,
  title = {Absolute Absorption on Rubidium {{D}} Lines: Comparison between Theory and Experiment},
  author = {Siddons, Paul and Adams, Charles S and Ge, Chang and Hughes, Ifan G},
  year = 2008,
  month = jul,
  journal = {Journal of Physics B: Atomic, Molecular and Optical Physics},
  volume = {41},
  number = {15},
  pages = {155004},
  issn = {0953-4075},
  doi = {10.1088/0953-4075/41/15/155004},
  urldate = {2025-10-30},
}

@inproceedings{hartl2004,
  title = {Carrier Envelope Phase Locking of an In-Line, Low-Noise {{Er}} Fiber System},
  booktitle = {Advanced {{Solid-State Photonics}} ({{TOPS}}) (2004), Paper 176},
  author = {Hartl, I. and Imeshev, G. and Cho, G. C. and Fermann, M. E. and Schibli, T. R. and Minoshima, K. and Onae, A. and Hong, F.-L. and Matsumoto, H. and Nicholson, J. W. and Yan, M. F.},
  year = 2004,
  month = feb,
  pages = {176},
  publisher = {Optica Publishing Group},
  doi = {10.1364/ASSP.2004.176},
  urldate = {2026-01-13},
  copyright = {\copyright{} 2004 Optical Society of America},
  langid = {english}
}

@article{jones2000,
  title = {Carrier-{{Envelope Phase Control}} of {{Femtosecond Mode-Locked Lasers}} and {{Direct Optical Frequency Synthesis}}},
  author = {Jones, David J. and Diddams, Scott A. and Ranka, Jinendra K. and Stentz, Andrew and Windeler, Robert S. and Hall, John L. and Cundiff, Steven T.},
  year = 2000,
  month = apr,
  journal = {Science},
  volume = {288},
  number = {5466},
  pages = {635--639},
  issn = {0036-8075, 1095-9203},
  doi = {10.1126/science.288.5466.635},
  urldate = {2026-01-13},
  langid = {english}
}

@article{RevModPhys_Safronova,
  title = {Search for new physics with atoms and molecules},
  author = {Safronova, M. S. and Budker, D. and DeMille, D. and Kimball, Derek F. Jackson and Derevianko, A. and Clark, Charles W.},
  journal = {Rev. Mod. Phys.},
  volume = {90},
  issue = {2},
  pages = {025008},
  numpages = {106},
  year = {2018},
  month = {Jun},
  publisher = {American Physical Society},
  doi = {10.1103/RevModPhys.90.025008},
  url = {https://link.aps.org/doi/10.1103/RevModPhys.90.025008}
}

@article{DCS_Budker,
  title = {Enhanced multichannel dual-comb spectroscopy of complex systems},
  author = {Aramyan, Razmik and Tretiak, Oleg and Sahoo, Sushree S. and Budker, Dmitry},
  journal = {Phys. Rev. Appl.},
  volume = {24},
  issue = {2},
  pages = {L021002},
  numpages = {7},
  year = {2025},
  month = {Aug},
  publisher = {American Physical Society},
  doi = {10.1103/7ktx-4h8m},
  url = {https://link.aps.org/doi/10.1103/7ktx-4h8m}
}

@article{Fortier:26,
author = {Tara M. Fortier and Andre N. Luiten and Helen S. Margolis},
journal = {Optica},
number = {1},
pages = {143--163},
publisher = {Optica Publishing Group},
title = {Optical atomic clocks: defining the future of time and frequency metrology},
volume = {13},
month = {Jan},
year = {2026},
url = {https://opg.optica.org/optica/abstract.cfm?URI=optica-13-1-143},
doi = {10.1364/OPTICA.575770},
}

@Article{ozone_spectroscopy,
AUTHOR = {Gorshelev, V. and Serdyuchenko, A. and Weber, M. and Chehade, W. and Burrows, J. P.},
TITLE = {High spectral resolution ozone absorption cross-sections{{-P}}art 1: Measurements, data analysis and comparison with previous measurements around 293 {{K}}},
JOURNAL = {Atmospheric Measurement Techniques},
VOLUME = {7},
YEAR = {2014},
NUMBER = {2},
PAGES = {609--624},
URL = {https://amt.copernicus.org/articles/7/609/2014/},
DOI = {10.5194/amt-7-609-2014}
}

@article{ozone_production,
author = {Tao, Madankui and Fiore, Arlene M. and Karambelas, Alexandra and Miller, Paul J. and Valin, Lukas C. and Judd, Laura M. and Tzortziou, Maria and Whitehill, Andrew and Teora, Amanda and Tian, Yuhong and Civerolo, Kevin L. and Tong, Daniel and Ma, Siqi and Adamo, Susana B. and Holloway, Tracey},
title = {Insights Into Summertime Surface Ozone Formation From Diurnal Variations in Formaldehyde and Nitrogen Dioxide Along a Transect Through New York City},
journal = {Journal of Geophysical Research: Atmospheres},
volume = {130},
number = {9},
year = {2025},
doi = {10.1029/2024JD040922}
}

\section{Supplementary Information}
\begin{figure*}[ht!]
\begin{centering}
\includegraphics[width=6 in]{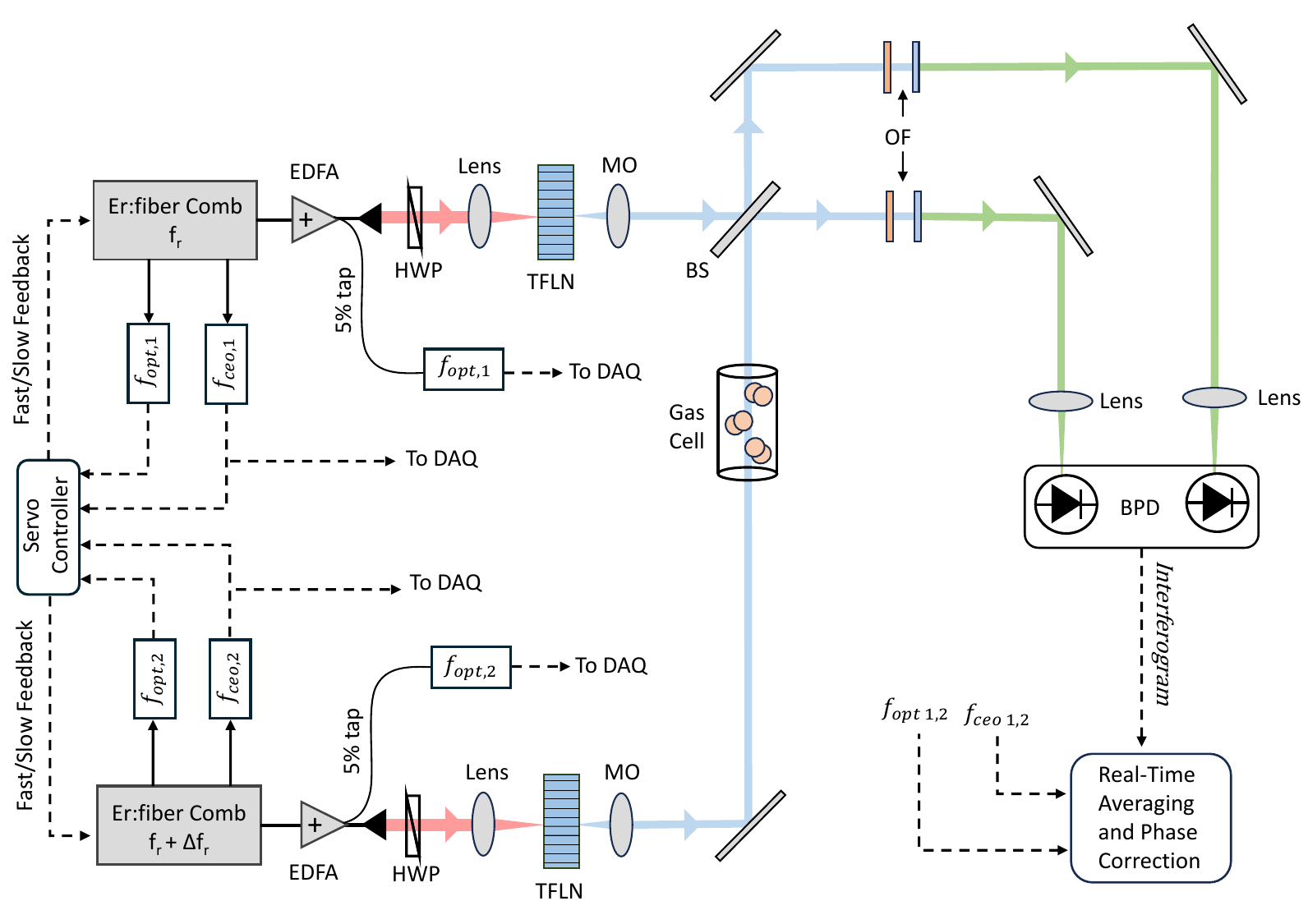}
\caption{Experimental setup and data acquisition details. The two combs are locked using out-of-loop optical beat detection and f-2f interferometry. A copy of the f$_{ceo}$ signal is sent to the data-acquisition system (DAQ). A second optical beat note is obtained using a five percent tap in the amplifiers and is sent to the DAQ. EDFA: Erbium-doped fiber amplifier, HWP: Half-wave plate, TFLN: Thin-film lithium niobate, MO: Microscope objective, BS: Beam-splitter, OF: Optical filter, BPD: Balanced photo-detector}
\label{fig:experimental_setup}
\end{centering} 
\end{figure*}
\subsection{Experimental Setup and Data Acquisition Details}
A detailed schematic of our experimental setup is shown in Supplemental Figure \ref{fig:experimental_setup}. One output-port of each comb is sent to the DCS system. The DCS system is based on a pair of Menlo frequency combs operating near 100~MHz repetition rates. A purpose-built and dispersion-managed erbium doped fiber amplifier was used to amplify the average power of our pulses to approximately 300~mW, producing sub-100~fs pulse durations. Approximately 2~m of polarization-maintaining gain fiber was pumped by three diode lasers operating near 980~nm: two back-pumping (1~W each) and one forward-pumping (750~mW). 
\par
The remaining two output-ports of the comb are used for stabilizing the repetition rate (f$_r$) and carrier-envelope offset (f$_{ceo}$). The optical beat for stabilizing the repetition rate is obtained via beating with a cavity-stabilized continuous-wave (CW) laser at 1550~nm. The carrier-envelope offset is measured using a standard f-2f interferomter\cite{jones2000, hartl2004}. Both signals are fed to servo controllers for PID feedback control with the oscillators. A portion of each f$_{ceo}$ RF signal is split off, amplified, and sent to the data acquisition system. A second pair of optical beat notes are obtained by beating comb light from a 5\% tap in the amplifier output with the same CW laser at 1550~nm, and are subsequently sent to the data acquisition system.
\par
The two TFLN waveguides used in this experiment share the same design but have slightly different geometries due to availability. Both waveguides were fabricated using a commercially available X-cut wafer with a 710~nm-thick lithium niobate layer on top of a $\sim$4.7~$\mu$m-thick silica buffer layer. Both waveguides have an etch-depth of 380~nm and width of 1800~nm, with a chirped poling region that linearly decreases the poling period from $\Lambda=12.5~\mu$m to $\Lambda=2.5~\mu$m in a length of 3.6~mm. They differ from one another in the length of their un-poled region at the beginning of the waveguide (0.9~mm and 3~mm), the waveguide cladding (SiO$_2$ and air), and their insertion loss (5-7~dB and 10~dB), respectively. In future experiments it is be desirable to have two identical waveguides to better enforce spectral overlap between the two combs and improve the SNR \cite{Newbury:sensitivity}. \par
\begin{figure*}[ht!]
\begin{centering}
\includegraphics[width=6 in]{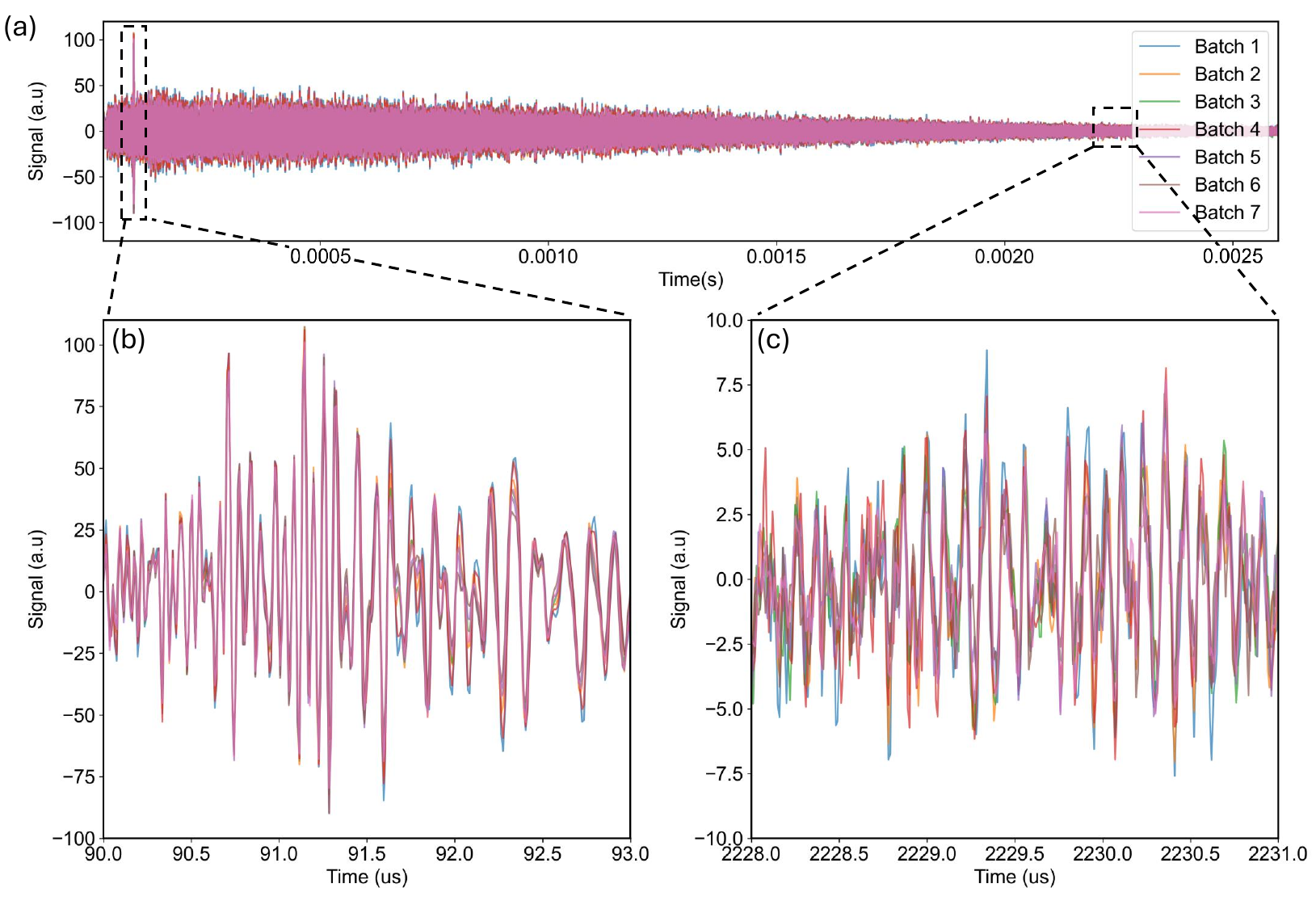}
\caption{Molecular free-induction decay of iodine. The full free induction decay is shown in the top panel. Zoom-in of the etalon near zero time-delay is shown in panel (b). Zoom-in of the FID at longer time-delay is shown in panel (c).}
\label{fig:7}
\end{centering} 
\end{figure*}
For the iodine spectroscopy experiments presented in the main text, the vapor cell had a length of $L=105$~mm and diameter of $D=20$~mm. Solid iodine is present inside the cell, and we estimate the iodine vapor pressure using Equation 1 from \cite{Reinhardt2007}.
\begin{equation}
    \log(p) = -\frac{3512.830}{T+273.15} - 2.013\times \log(T+273.15) + 18.37971
\end{equation}
where {\it T} is temperature in degrees Celsius and {\it p} is the iodine vapor pressure in Pascal. For room temperature of $T=25$~C, the vapor pressure of iodine in the cell is approximately 41.4~Pascal or 310~mTorr.\par

The signal processing used in the data acquisition system is described in-depth in Ref.\cite{WalshGPU}, and is available commercially (www.elanspectral.com) but will be described briefly here. A fast-correction algorithm uses the supplied references ($f_{ceo1,2}, f_{opt1,2}$) to extract phase-noise information. The four referencing signals are phase demodulated and combined such as to first retrieve the phase evolutions $\delta\phi_{ceo}$ and $\delta\phi_{opt}$. Knowing that $\delta\phi_{opt}=n \delta\phi_{fr} + \delta\phi_{ceo}$, one can then obtain $n \phi_{fr}$.
\par
The phase evolution of $f_r$ and $f_{ceo}$ are combined in a way such as to perfectly cancel the phase noise of any selected tooth in the electrical comb, including taking into account any non-linear transformation of the interferogram light. For instance, to correct for the k-th mode in the third harmonic spectrum, one would compute $\phi_{mode}=3(k\phi_{fr}/n + \phi_{ceo})$. Applying $e^{-\phi_{mode}}$ to the complex analytic interferogram brings the selected mode exactly at zero in electrical frequency and the signal can thereafter be resampled using $n\phi_{fr}$.
\par
These steps are performed once using the referencing beat notes to correct for fast fluctuations and a second time afterwards using phase timing information inferred from the interferogram centerbursts (i.e. self-correction) to remove any slow residual out-of-loop fluctuations. At the second resampling step it is ensured that the same grid of optical delay, having an integer number of points per interferograms, is always used.
\par
To demonstrate the effectiveness of the implemented real-time phase-correction, we plot the measured free-induction decay of molecular iodine in each of our 1000~s averages alongside one another in Supplemental Fig. \ref{fig:7}(a). In Supplemental Fig. \ref{fig:7}(b) we zoom-in on the etalon present in the FID near the centerburst, showing the phases of consecutive batches of interferograms are very well aligned over the course of over an hour. Moving much further from the centerburst, Fig. \ref{fig:7}(c) shows the phases of the interferogram batches remain well-aligned even at large time-delays.\par

\subsection{Rubidium/Sodium Spectroscopy and Modeling}
We extract the center-of-mass frequencies of the rubidium D lines by constructing two models similar to Ref. \cite{Siddons2008}. The equation representing the absorption profile of each individual hyperfine transition is a Voigt profile ($V$) with Lorentzian and Gaussian linewidths determined by the natural ($\Gamma$) and Doppler ($\sigma$) linewidths, respectively. Specifically, the absorption profile for a single transition is,
\begin{equation}
    \alpha_{F_g F_e}(\Delta) = A C_F^2 n\frac{1}{2 I+1}  V(\nu_{CM} - \Delta;\Gamma, \sigma),
\end{equation}
where $C_F^2$ is the relative transition strength, $n$ is the relative abundance of the isotope, $I$ is the degeneracy of the isotope's ground state, $\Delta$ is the relative frequency of the transition, and $A$ is an arbitrary scale factor shared by all transitions in each measurement. The transmission spectrum of each line is a superposition of each hyperfine transition, such that,
\begin{equation}
    T = \exp[-\sum \alpha_i].
\end{equation}
The D2 and D1 sodium models use the same formula as rubidium, but use corresponding numbers from Ref. \cite{steckSodium}.

\begin{table*}[ht!]
    \centering
    \begin{tabular}{|l|l|l|}
    \hline
         Line & $\nu_{CM}$ (GHz) & Offset (MHz) \\
         \hline
         Rb D2 & 384,230.436(3) & 9\cite{Siddons2008} \\
         Rb D1 & 377,107.409(7) & 2\cite{Siddons2008} \\
         Na D2 & 508,848.718(16) & 2\cite{juncar1981}\\
         Na D1 & 508,333.152(25) & -43\cite{juncar1981}\\
         \hline
    \end{tabular}
    \caption{Fitted center-of-mass frequencies for the rubidium/sodium D2/D1 lines. Included are comparisons to other absolute frequency measurements. The offset is defined as the difference between our measurements and those from the literature, such that a positive sign implies that our present measurement is higher in frequency than the literature value.}
    \label{tab:linecenters}
\end{table*}

We independently fit each of the four models to our data over  
$\sim$10 GHz spectral ranges. This keeps low-frequency fluctuations in the background level (due to an etalon) from significantly affecting the fit statistics. Each model has three free parameters: the amplitude scale factor $A$, the center-of-mass frequency $\nu_{CM}$, and a vertical offset to account for the imperfect background level. From this fit, we derive each line's center-of-mass frequency and corresponding statistical uncertainty. These are reported in Table \ref{tab:linecenters}.

\subsubsection{Absolute Frequency Uncertainty}
We estimate the frequency uncertainty of our spectrometer from a statistical uncertainty in the repetition rate of each comb. The comb equation $\delta f_n = n\delta f_r + \delta f_{0}$ tells us how uncertainties in the repetition rate $f_r$ and carrier-envelope offset $f_0$ limit the knowledge of the frequency of a particular optical comb-tooth $f_n$. Each repetition rate is measured on a counter over the duration of the measurement and the uncertainty is estimated from the Allan deviation. All frequencies of the combs and data acquisition are referenced to a GPS-steered oscillator which has fractional uncertainty below $1 \times 10^{-11}$. For our optically-stabilized and self-referenced frequency combs, the uncertainty in $f_0$ is additive and negligible. The optical stabilization provides excellent short term stability, so the variance in $f_r$ is dominated by the reference oscillator for averages under $\sim$1 minute. For the longer averages performed in this paper, $\delta f_r$ is dominated by cavity drift. For our spectrometer operating around 550~THz for 2 hours, as in our iodine measurements, we estimate a frequency uncertainty of 10~kHz.

\subsubsection{Gas Cells and Spectroscopy}
The rubidium cell contains a natural isotope ratio of $^{85}$Rb ($\approx$78\%) and $^{87}$Rb ($\approx$22\%). The vapor pressure of $^{85}$Rb and $^{87}$Rb in the solid-phase is estimated by Equation 2 from Refs\cite{steckRubidium85,steckRubidium87},
\begin{equation}
    \log(p) = 2.881 + 4.857 - \frac{4215}{T}
\end{equation}
where {\it T} is the temperature in Kelvin and {\it p} is the pressure in Torr. In the main text, the results shown in Fig.\ref{fig:5}(c)-(d) were taken at two different temperatures. For the D1 measurements (Fig.\ref{fig:5}(a),(c)) the vapor cell was heated to approximately T=310~K, corresponding to a vapor pressure of $\sim$1.8~$\mu$Torr, and interferograms were averaged for 30 minutes. The vapor cell was kept at room temperature $T=25$~C for the D2 measurements (Fig.\ref{fig:5}(d)), corresponding to a vapor pressure of $\sim$0.4~$\mu$Torr. Interferograms were averaged for 30 minutes during D2 measurements.


\par

We perform sodium spectroscopy above the melting point of solid sodium. The vapor pressure of sodium in the liquid-phase is estimated by Equation 4, taken from \cite{steckSodium},
\begin{equation}
    \log(p) = 2.881 + 4.704 - \frac{5377}{T}
\end{equation}
where {\it T} is the temperature in Kelvin and {\it p} is the pressure in Torr. For our experiments at $T=429$~K, the vapor pressure of sodium in the cell is approximately 11.2~$\mu$Torr

\subsection*{Dual-Comb Spectroscopy of Nitrogen Dioxide}
\begin{figure*}[ht!]
\centering
\includegraphics[width=18cm]{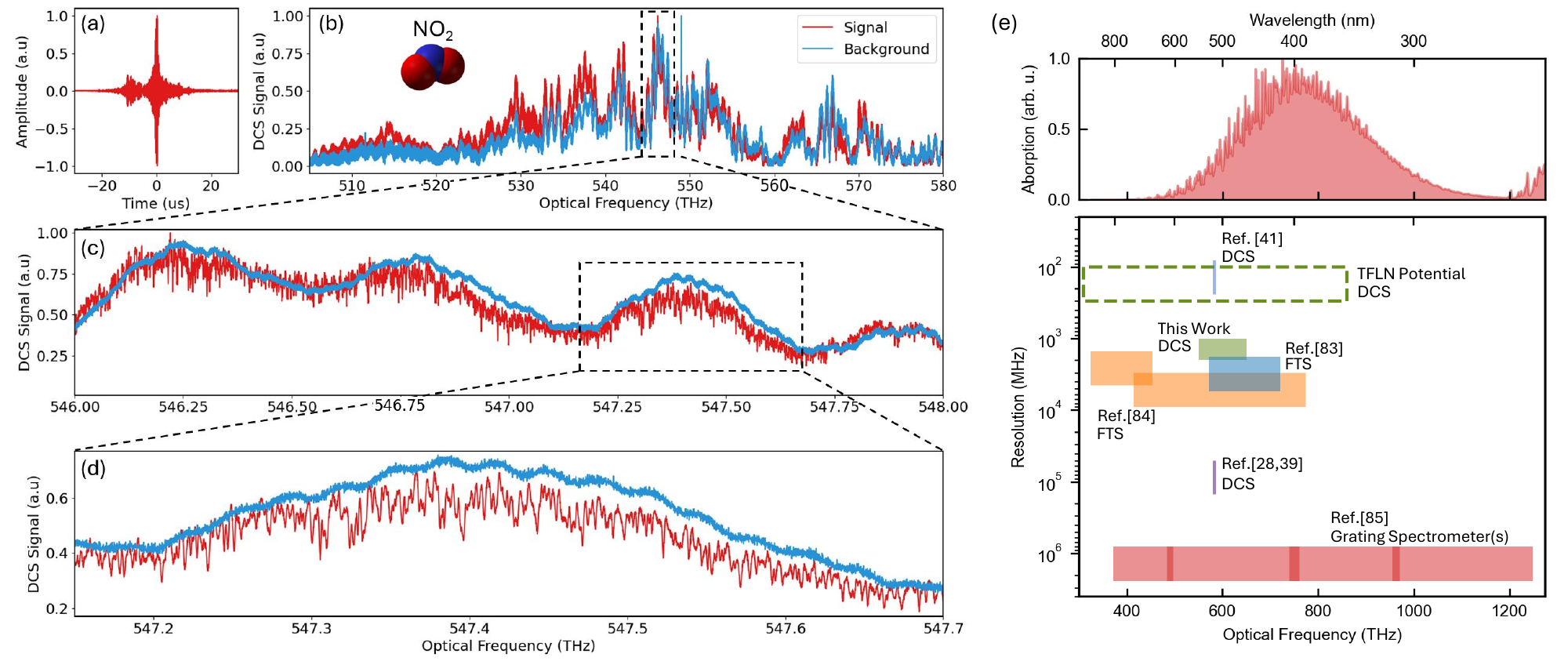}
\caption{Dual-comb interferogram and spectrum from NO\textsubscript{2} spectroscopy. (a) Time-averaged interferogram. (b) Corresponding dual-comb spectrum (red) and measured background dual-comb spectrum (blue). (c) Expansion of region in (b). (d) Expansion of region in (c), showing the high-density absorption features of nitrogen dioxide. (e) Comparison of spectral resolutions and bandwidths afforded by different spectroscopic instrumentation applied to measuring nitrogen dioxide. The UV-VIS absorption spectrum of NO$_2$ is shown in the top panel for reference. The green dashed box indicates the full bandwidth covered by our frequency combs.}
\label{fig:8}
\end{figure*}

A preliminary experiment highlights the potential of our broadband visible DCS system for the interpretation of the complicated nitrogen dioxide (NO$_2$) spectrum. We measure features over approximately 90~THz of optical bandwidth with an estimated resolution of 1~GHz.\par

The gas cell used for nitrogen dioxide experiments was purpose built with a length of $L=160$~mm and diameter $D=17$~mm. The empty cell pressure was approximately 30~mTorr before being filled to a pressure of 650~Torr with 99.9\% purity nitrogen dioxide. The cell was evacuated to a pressure of 100~mTorr before obtaining a background spectrum.

Supplemental Fig. \ref{fig:8}(a) shows the measured time-domain interferogram resulting from approximately 2.5 hours of coherent averaging. The corresponding dual-comb spectrum is shown in Supplemental Fig. \ref{fig:8}(b) along with the background spectrum taken once the NO$_2$ was evacuated from the gas cell. Apparent is the fact that the dual-comb spectrum changed over the span of 2.5 hours and therefore the background spectrum falls below the signal spectrum in many areas. This is most likely a result of drifts in the coupling into and out of the TFLN waveguides (e.g. optomechanical drift and vibrations and lack of temperature control of the waveguides).
\par
An expanded region of the DCS spectrum where the spectral drift is minimal (Supplemental Fig. \ref{fig:8}(d)) provides a useful local comparison between the background and signal spectra. Here we observe a high-density of narrow-linewidth transitions with complicated structure. In the 0.5~THz span shown, we estimate to have measured almost 150 molecular absorption features, demonstrating the unique capabilities of this system in interrogating complex absorption spectra over broad optical bandwidths.

In Supplemental Fig. \ref{fig:8}(e) we plot the bandwidth and resolution of different spectroscopic techniques that have been applied to measuring nitrogen dioxide\cite{Nizkorodov2004,Vandaele2002,BOGUMIL2003,Pal2025,Eber:2024}. The UV-VIS absorption band of NO$_2$ measured with a grating spectrometer at a spectral resolution on the order of hundreds of gigahertz is plotted above to indicate the typical span of molecular absorption bands in this spectral region\cite{BOGUMIL2003}. Our system measured one of the highest resolution spectra of this molecule, but the bandwidth is still less than that achieved by FTS. For comparison, the dashed box near our work indicates the full bandwidth produced by the TFLN waveguides with comb-mode resolution (100~MHz).  Other approaches, such as harmonic generation of few-cycle pulses in high-bandgap materials, could provide an avenue to broadband coverage at wavelengths below 350~nm\cite{Lesko2022}.\par

\subsection{SNR Analysis}
We use the background and signal dual-comb spectra for the NO$_2$ experiments to estimate the SNR across the dual-comb spectrum. The noise is approximated with a moving standard deviation across the background spectrum, using bins of 1~THz of bandwidth separated by steps of 10~GHz. The calculated standard deviations are then used to calculate the SNR of our dual-comb spectrum as a function of frequency. The bandwidth of our dual-comb experiment is then estimated as the bandwidth within which the dual-comb SNR remains greater than unity. This analysis gives a DCS bandwidth of approximately 90~THz in the nitrogen dioxide experiments presented.

A similar analysis was carried out with our dual-comb spectrum of I$_2$. A 1~THz bandwidth at the low frequency edge of our dual-comb spectrum (near 481~THz) with no signal was used to calculate a standard deviation to approximate the noise. Then, this standard deviation was used to estimate our dual-comb SNR as a function of frequency. We estimate our dual-comb bandwidth as the bandwidth within which the dual-comb SNR remains greater than unity. For the iodine experiments presented, the DCS bandwidth is approximately 120~THz.

\end{document}